\documentclass[gmd, manuscript]{copernicus}

\begin{document}

\title{Thetis coastal ocean model: discontinuous Galerkin discretization for the three-dimensional hydrostatic equations}


\Author[1, $\dagger$]{Tuomas}{K\"{a}rn\"{a}}
\Author[3]{Stephan C.}{Kramer}
\Author[2,4,*]{Lawrence}{Mitchell}
\Author[2]{David A.}{Ham}
\Author[3]{Matthew D.}{Piggott}
\Author[1]{Ant\'onio M.}{Baptista}

\affil[1]{Center for Coastal Margin Observation \& Prediction, Oregon Health \& Science University, Portland, Oregon, USA}
\affil[2]{Department of Mathematics, Imperial College London, London, United Kingdom}
\affil[3]{Department of Earth Science and Engineering, Imperial College London, London, United Kingdom}
\affil[4]{Department of Computing, Imperial College London, London, United Kingdom}
\affil[$\dagger$]{Present address: Finnish Meteorological Institute, Helsinki, Finland}
\affil[*]{Present address: Department of Computer Science, Durham
  University, Durham, United Kingdom}

\runningtitle{Thetis: discontinuous Galerkin discretization}

\runningauthor{T. K\"{a}rn\"{a} et al.}

\correspondence{tuomas.karna@gmail.com}

\received{}
\pubdiscuss{} 
\revised{}
\accepted{}
\published{}


\nolinenumbers

\firstpage{1}

\maketitle

\begin{abstract}
Unstructured grid ocean models are advantageous for simulating the coastal ocean and river-estuary-plume systems.
However, unstructured grid models tend to be diffusive and/or computationally expensive which limits their applicability to real life problems.
In this paper, we describe a novel discontinuous Galerkin (DG) finite element discretization for the hydrostatic equations.
The formulation is fully conservative and second-order accurate in space and time.
Monotonicity of the advection scheme is ensured by using a strong stability preserving time integration method and slope limiters.
Compared to previous DG models advantages include a more accurate mode splitting method, revised viscosity formulation, and new second-order time integration scheme.
We demonstrate that the model is capable of simulating baroclinic flows in the eddying regime with a suite of test cases.
Numerical dissipation is well-controlled, being comparable or lower than in existing state-of-the-art structured grid models.
\end{abstract}

\introduction  

Numerical modeling of the coastal ocean is important for many environmental and industrial applications.
Typical scenarios include modeling circulation at regional scales, coupled river-estuary-plume systems, river networks, lagoons, and harbors.
Length scales range from some tens of meters in rivers and embayments to tens of kilometers in the coastal ocean; water depth ranges from less than a meter to kilometer scale at the shelf break.
The time scales of the relevant processes range from minutes to hours, yet typical simulations span weeks or even decades.
The dynamics are highly non-linear, characterized by local small-scale features such as fronts and density gradients, internal waves, and baroclinic eddies.
These physical characteristics imply that coastal ocean modeling is intrinsically multi-scale, which imposes several technical challenges.

Most coastal ocean models solve the hydrostatic Navier-Stokes equations under the Boussinesq approximation -- a valid approximation for mesoscale and sub-mesoscale (1 \unit{km}) processes.
Small-scale processes (< 100 m) are, however, inherently three-dimensional where non-hydrostatic effects can be important, especially in areas with pronounced density structure and stratification \citep{marshall1997b,mahadevan2006}.
Non-hydrostatic modeling requires very high horizontal mesh resolution, which is currently only feasible in relatively small sub-regions (e.g. at the mouth of an estuary; \citealt{shi2016}) due to its high computational cost.

Historically, regional ocean models have used structured, (deformed) rectilinear lattice grids.
Although structured grids offer better computational performance \citep{danilov2008,danilov2013}, unstructured grids are generally preferred in coastal domains as they can better represent the complex coastal topography and local features \citep{deleersnijder2008a,danilov2013,piggott2013}.
Due to the large geometrical aspect ratio of the oceans (length versus depth), most models utilize computational grids that are layered in the vertical direction.
Typical approaches include the terrain-following sigma levels \citep{blumberg1987}, equipotential $z$ levels \citep{griffies2005}, isopycnal coordinates \citep{bleck1978}, and their generalizations \citep[e.g.][]{song1994,bleck2002}.

In this article, we focus on solving the hydrostatic equations on an unstructured grid.
While many unstructured grid models exist, their drawbacks tend to be excessive numerical diffusion that smooths out important physical features \citep{karna2015,karna2016a,ralston2017a}, and/or high computational cost.
To address these issues, we propose a novel finite element solver for the hydrostatic equations, based on discontinuous Galerkin discretization methods.

Maintaining high numerical accuracy is crucial in ocean applications.
The ocean is a forced dissipative system where the mixing of water masses only takes place at the molecular level \citep{griffies2004a}.
In practice, however, the finite grid resolution and numerical schemes used by the model introduce mixing rates of tracers and momentum that can be orders of magnitude larger than physical mixing \citep{burchard2008b,rennau2009,hiester2014}.
Such spurious, numerical mixing is often dominated by the discretization of advection \citep{marchesiello2009,griffies2000c}, but it can arise from other components as well, such as (implicit) time integration methods \citep{shchepetkin2005}, or various filters introduced to improve numerical stability \citep{danilov2012,zhang2016}.
In addition, wetting and drying schemes may introduce additional dissipation in order to stabilize the barotropic equation in the drying regime.
We reserve consideration of this important latter topic for a future publication.

In global circulation models, numerical mixing is a major bottleneck as (diapycnal) diffusion is very low in the deep ocean basins and water masses can remain largely unchanged for hundreds of years  \citep{griffies2004a,griffies2000c}.
Numerical mixing can, however, be a major issue in coastal domains as well:
coastal oceans are characterized by strong density gradients, fronts between water masses (e.g. in river plumes), small-scale dynamics (e.g. internal waves and hydraulic jumps), and baroclinic eddies.
An overly diffusive model can, therefore, fail to capture many essential physical features of these domains:
it can smear out fronts, underestimate the intrusion of saline waters into embayments \citep{burchard2008b, hofmeister2010, karna2015, ralston2017a}, or misrepresent mixing in river plumes.

The most common spatial discretization scheme is the finite volume (FV) method, used in MITgcm \citep{marshall1997}, GETM \citep{burchard2002b}, ROMS \citep{shchepetkin2003,shchepetkin2005}, MPAS-Ocean \citep{ringler2013,petersen2015}, UnTRIM \citep{casulli2000}, FVCOM \citep{chen2003}, SUNTANS \citep{fringer2006}, FESOM2 \citep{danilov2016}, and others.
The FV method is well suited for advection-dominated problems, provides strict conservation of volume and mass, and yields good computational performance. FV methods are nominally only first-order accurate, but higher-order approximations can be introduced by increasing the size of the numerical stencil (e.g. in high-order advection schemes, \citealt{shchepetkin1998}).

Some unstructured grid models are based on the continuous Galerkin Finite Element (FE) method or hybrid FE-FV formulations. 
Such models include ADCIRC \citep{luettich2004}, SELFE \citep{zhang2008}, and SCHISM \citep{zhang2016}, and the earlier version of FESOM \citep{wang2014}.
The continuous FE method is ideal for solving elliptic equations but requires stabilization for advection \citep[see][and references therein]{wang2008}.
In addition, these methods involve solving a fully-coupled global system which is less efficient in parallel applications compared to the FV method \citep{danilov2012,danilov2016}.

In recent years, discontinuous Galerkin (DG) methods have gained attention in geophysical modeling \citep{dawson2005,aizinger2007,blaise2010,comblen2010a,karna2012,karna2013}.
DG discretization resembles the FV method because it is local (i.e. elements are only connected by inter-element fluxes), fully conservative, and well-suited for advective problems, yet it offers higher-order accuracy.
This article presents a DG discretization for the hydrostatic equations.
Our goal is to design an efficient unstructured grid solver where numerical accuracy is not compromised.
Specifically, we aim to meet the following design criteria:

\begin{itemize}
 \item a vertically extruded, layered mesh;
 \item accurate representation of free surface dynamics;
 \item second-order accurate, monotone tracer advection scheme;
 \item explicit time integration of 3D variables (except for vertical diffusion);
 \item and low numerical mixing.
\end{itemize}

Based on the advection scheme requirements, we have chosen to use linear discontinuous Galerkin elements for tracers, combined with a slope limiter \citep{kuzmin2010} and a strong stability preserving (SSP) time integration scheme \citep{shu1988b,shu1988,gottlieb1998,gottlieb2005,gottlieb2009}.
This choice ensures that the scheme is second-order in smooth areas,
while slope limiting combined with the SSP time integration scheme ensure monotonicity (i.e. no overshoots).
The movement of the free surface is taken into account with an arbitrary Lagrangian-Eulerian (ALE) formulation \citep{donea2004}, where the mesh moves in the vertical direction.
The ALE formulation guarantees strict local and global conservation of volume and tracers and allows for the use of generic vertical grids \citep{petersen2015}.

All numerical ocean models include some form of friction, either in the form of a numerical closure or a physical parametrization \citep{griffies2000b}.
Numerical closure involves adding a sufficient amount of dissipation to maintain numerical stability.
There is a wealth of literature about stable finite volume \citep[e.g.,][]{danilov2012} and finite element discretizations \citep[e.g.,][]{hanert2003,cotter2009,cotter2009b,comblen2010,mcrae2014} for rotational shallow water equations.
Most of these schemes are stable for external gravity waves, and hence do not require any additional dissipation.
Solving the 3D hydrostatic equations under strong baroclinic forcing, however, generates noise at the grid-scale that does require dampening.
A common approach is to add some form of viscosity proportional to the grid Reynolds number \citep{griffies2000b,ilicak2012}.
\cite{griffies2000b} argue that conventional Laplacian viscosity has too wide a spectrum and tends to dissipate physically relevant (larger) scales too much.
They show that biharmonic viscosity dissipates smaller scales more, and is thus more appropriate for removing noise at the grid-scale.
In contrast to numerical closures, physical parametrizations aim to represent unresolved sub-grid-scale processes, such as strong lateral mixing near coasts or mixing at the bottom boundary layers.
In this article, we focus on numerical closures; the presented viscosity schemes are mostly motivated by numerical stability considerations.

In this article, we present an efficient DG implementation of the three-dimensional hydrostatic equations.
The model is implemented in the \emph{Thetis} project -- an open source coastal ocean circulation model freely available online (see \href{http://thetisproject.org}{thetisproject.org}).
Thetis implements both a 2D depth-averaged circulation model and a full 3D hydrostatic model, the latter of which is discussed herein.

Thetis is implemented using the Firedrake finite element modeling platform \citep[\url{www.firedrakeproject.org};][]{rathgeber2016}.
We have chosen Firedrake because of its flexibility, and support for extruded meshes \citep{mcrae2016,bercea2016}.
Firedrake uses high-level abstractions for describing the weak
formulation of partial differential equations, specifically the
Unified Form Language \citep{alnaes2014}, and automated code
generation to produce efficient C code \citep{homolya2017,luporini2017} and just-in-time compilation.
As such it is an extremely flexible modeling framework that does not sacrifice computational efficiency; it is also an ideal platform for experimenting and benchmarking different discretizations.
Automated code generation can also support different target hardware architectures, making it attractive for current and emerging high-performance computing platforms.
In addition, Firedrake can automatically derive the adjoint of the forward model \citep{farrell2013}, permitting inverse modeling applications such as parameter optimization and data assimilation.

The governing equations are presented in Section \ref{sec:equations}, followed by their DG finite element discretization in Section \ref{sec:fe_discretization}.
The second-order coupled time integration scheme is described in Section \ref{sec:timeintegration}. Numerical tests are presented in Section \ref{sec:testcases}.

\section{Governing equations}\label{sec:equations}

Let $\Omega$ be the three-dimensional domain that spans from the sea floor $z=-h(x,y)$ to the free surface $z=\eta(x,y)$; the bottom and top surfaces are denoted by $\Gamma_b$ and $\Gamma_s$, respectively. Total water column depth is thus $H = \eta + h$.
The two-dimensional horizontal domain is denoted by $\Gamma_0$.

The horizontal momentum equation reads

\begin{align}
\begin{split}
\frac{\partial \mathbf{u}}{\partial t} + \boldsymbol{\nabla}_h \cdot (\mathbf{u} \mathbf{u}) 
+ \frac{\partial \left(w\mathbf{u} \right)}{\partial z} + f\mathbf{e}_z\wedge\mathbf{u} + \frac{1}{\rho_0}\boldsymbol{\nabla}_h p\\
 =
\boldsymbol{\nabla}_h \cdot \left( \nu_h \boldsymbol{\nabla}_h \mathbf{u} \right) + \frac{\partial}{\partial z}\left( \nu  \frac{\partial \mathbf{u}}{\partial z}\right),
\label{eq:mom_original}
\end{split}
\end{align}
where $\mathbf{u} = (u, v)$ and $w$ denote the horizontal and vertical velocity, respectively; $\boldsymbol{\nabla}_h$ is the horizontal gradient operator; $\wedge$ denotes the cross product operator;
$f$ is the Coriolis parameter; $\mathbf{e}_z$ is the vertical unit vector; $p$ is the pressure; and $\nu_h$ and $\nu$ are the horizontal and vertical diffusivity, respectively.
Water density is defined as $\rho = \rho_0 + \rho'(T, S, p)$, where $T,S$ stand for temperature and salinity, respectively, and $\rho_0$ is a constant reference density.

Under the hydrostatic assumption
the horizontal pressure gradient can be written as a combination of external, internal, and atmospheric pressure gradients:

\begin{align}
 \frac{1}{\rho_0}\boldsymbol{\nabla}_h p = g\boldsymbol{\nabla}_h \eta + g\boldsymbol{\nabla}_h r + \frac{1}{\rho_0}\boldsymbol{\nabla}_h p_{\text{atm}},
\end{align}
where $p_{\text{atm}}$ is the atmospheric pressure acting on the sea surface, and

\begin{align}
 r = \frac{1}{\rho_0}\int_{z}^\eta \rho' dz' \label{eq:baroc_head}
\end{align}
is the baroclinic head.
For brevity the internal pressure gradient field is denoted as $\mathbf{F}_{\text{pg}} = g\boldsymbol{\nabla}_h r$.

Neglecting atmospheric pressure, the full horizontal momentum equation reads

\begin{align}
 \frac{\partial \mathbf{u}}{\partial t} + \boldsymbol{\nabla}_h \cdot (\mathbf{u} \mathbf{u}) + \frac{\partial \left(w\mathbf{u} \right)}{\partial z} +
 f\mathbf{e}_z\wedge\mathbf{u} + g\boldsymbol{\nabla}_h \eta + \mathbf{F}_{\text{pg}}
 =
\boldsymbol{\nabla}_h \cdot \left( \nu_h \boldsymbol{\nabla}_h \mathbf{u} \right)  + \frac{\partial}{\partial z}\left( \nu  \frac{\partial \mathbf{u}}{\partial z}\right). \label{eq:mom}
\end{align}
Vertical velocity $w$ is diagnosed from the continuity equation:

\begin{align}
\boldsymbol{\nabla}_h \cdot \mathbf{u}  + \frac{\partial w}{\partial z} = 0.  \label{eq:continuity3d}
\end{align}
Water temperature and salinity are modeled with an advection-diffusion equation of the form

\begin{align}
 \frac{\partial T}{\partial t} + \boldsymbol{\nabla}_h \cdot (\mathbf{u} T) + \frac{\partial \left(w T \right)}{\partial z} =
\boldsymbol{\nabla}_h \cdot \left( \mu_h \boldsymbol{\nabla}_h T \right)  + \frac{\partial}{\partial z}\left( \mu \frac{\partial T}{\partial z}\right), \label{eq:tracer}
\end{align}
where $\mu_h,\mu$ stand for the horizontal and vertical (eddy) diffusivity, respectively.

At the bottom boundary we impose quadratic bottom stress

\begin{align}
\left( \nu_h \mathbf{n}_h \cdot \boldsymbol{\nabla}_h \mathbf{u} + \nu n_z \frac{\partial \mathbf{u}}{\partial z} \right)\bigg\vert_{\mathbf{x}\in\Gamma_b} &= \frac{\boldsymbol{\tau}_b}{\rho_0}, \\
\frac{\boldsymbol{\tau}_b}{\rho_0} &= C_d | \mathbf{u}_{bf} | \mathbf{u}_{bf}, \label{eq:bottom_stress}
\end{align}
where $C_d$ is the drag coefficient, and $\mathbf{u}_{bf}$ is the velocity in the middle of the bottommost element.
$\mathbf{n}=(n_x, n_y, n_z)$ is the outward normal vector, and $\mathbf{n}_h=(n_x,n_y,0)$ its horizontal projection.
The bottom boundary condition is treated implicitly; \eqref{eq:bottom_stress} is linearized by keeping the magnitude $| \mathbf{u}_{bf} |$ fixed at the ``old'' value while solving for $\mathbf{u}$ (and $\mathbf{u}_{bf}$).
Typically $C_d$ is computed from the logarithmic law of the wall \citep[e.g.][]{karna2013}.

\subsection{Mode splitting}\label{sec:modesplitting}

Following \cite{higdon1997} we split the horizontal velocity field into depth-averaged $\bar{\mathbf{u}}$ and deviation $\mathbf{u}' = \mathbf{u} - \bar{\mathbf{u}}$ components.
The depth-averaged momentum equation is then defined as

\begin{align}
 \frac{\partial \bar{\mathbf{u}}}{\partial t} + f\mathbf{e}_z\wedge\bar{\mathbf{u}} + g\boldsymbol{\nabla}_h \eta = \mathbf{G}, \label{eq:mom2d}
\end{align}
where $\mathbf{G}$ is a forcing term used to couple the 2D and 3D modes.
This equation is complemented with the depth-averaged continuity (free surface) equation:

\begin{align}
\frac{\partial \eta}{\partial t} + \boldsymbol{\nabla}_h\cdot\left(H\bar{\mathbf{u}}\right) = 0. \label{eq:freesurface}
\end{align}

The 2D system \eqref{eq:mom2d}--\eqref{eq:freesurface} contains the fast-propagating, rotational surface gravity waves.
The corresponding equation for $\mathbf{u}'$ is obtained by subtracting \eqref{eq:mom2d} from \eqref{eq:mom} \citep{higdon1997}:

\begin{align}
\begin{split}
 \frac{\partial \mathbf{u}'}{\partial t} + \boldsymbol{\nabla}_h \cdot (\mathbf{u} \mathbf{u}) + \frac{\partial \left(w\mathbf{u} \right)}{\partial z} +
 f\mathbf{e}_z\wedge\mathbf{u}' + \mathbf{F}_{\text{pg}}\\
 =
\boldsymbol{\nabla}_h \cdot \left( \nu_h \boldsymbol{\nabla}_h \mathbf{u} \right) + \frac{\partial}{\partial z}\left( \nu  \frac{\partial \mathbf{u}}{\partial z}\right) - \mathbf{G}. \label{eq:mom3d}
\end{split}
\end{align}

Note that the advection and viscosity terms are included in \eqref{eq:mom3d} without splitting, based on the assumption that these processes are slow enough to be captured with long time steps.
The Coriolis term, on the other hand, only contains the slow modes.
The vertical velocity $w$ only appears in the advection term, which is not split, and thus there is no need to split $w$.

\subsection{Coupling 2D and 3D modes}\label{sec:mode_coupling}

The 2D and 3D modes are coupled using the additional term $\mathbf{G}$ \citep{higdon1997,ringler2013}.
First, the 3D momentum equation \eqref{eq:mom3d} is solved with $\mathbf{G}=0$,  resulting in a velocity field $\mathbf{u}'$ that has a non-zero depth-average, generated by the advection and viscosity terms (that depend on $\bar{\mathbf{u}}$).
We then compute the depth-average $\overline{\mathbf{u}'}$ and apply a correction:
\begin{align}
 \mathbf{G} &= \overline{\mathbf{u}'}/\Delta t, \label{eq:uv_coupling} \\
 \mathbf{u}' &\leftarrow \mathbf{u}' - \mathbf{G} \Delta t \label{eq:uv_correction}
\end{align}
 to enforce zero depth-average.
By definition, the field $\mathbf{G}$ is a constant over the vertical, and it will be used as a forcing term in the 2D momentum equation \eqref{eq:mom2d} in the subsequent solve.
This procedure ensures that equations \eqref{eq:mom2d} and \eqref{eq:mom3d} sum up to \eqref{eq:mom} and $\int \mathbf{u}' dz = 0$.

\subsection{Equation of state}\label{sec:eq_of_state}

In this paper a linear equation of state is used:
\begin{align}
 \rho(T, S) = \rho_0 - \alpha_T (T-T_0) + \beta_S (S - S_0), \label{eq:lin_eos}
\end{align}
where $\alpha_T,\beta_S$ are the thermal expansion and saline contraction coefficients, respectively, and $T_0, S_0$ are reference temperature and salinity.
In all the test cases presented herein salinity does not contribute to water density ($\beta_S=0$).
Thetis also implements a full non-linear equation of state \citep{jackett2006}.

\subsection{Viscosity and turbulence closure}\label{sec:turbulence}

Baroclinic flows require some form of viscosity to filter out grid-scale noise.
In this paper we only consider Laplacian horizontal viscosity, set to a constant $\nu_h = U \Delta x / \text{Re}_h$ corresponding to the velocity scale $U$, horizontal mesh resolution $\Delta x$, and the desired grid Reynolds number $\text{Re}_h$. Here the velocity scale $U$ is taken as a global constant specific to each test case.
Unless otherwise specified, the horizontal diffusivity of tracers is zero.

In most test cases vertical viscosity is set to a constant.
In certain cases we use the gradient Richardson number dependent parametrization by \cite{pacanowski1981}:

\begin{equation}
\begin{aligned}
 \nu &= \frac{\nu_0}{(1 + \alpha \text{Ri})^n} + \nu_b, \\
 \mu &= \frac{\nu}{1 + \alpha \text{Ri}} + \mu_b, \label{eq:pacanowski}
\end{aligned}
\end{equation}

where $\text{Ri} = N^2/M^2$ is the gradient Richardson number, $N$ is the buoyancy frequency, and $M$ is the vertical shear frequency.
The background values are set to $\nu_b=\mu_b=2\times10^{-5}$ \unit{m^2\ s^{-1}}, while maximum viscosity is set to $\nu_0 = 2\times10^{-2}$ \unit{m^2\ s^{-1}};
the dimensionless parameters are $\alpha=10$ and $n=2$ \citep{wang2008a}.
More sophisticated turbulence closures will be addressed in future work.

\section{Finite element discretization}\label{sec:fe_discretization}

This section describes the spatial discretization of the governing equations.
In Section \ref{sec:function_spaces} we define the finite element function spaces, followed by the weak forms of the underlying equations.

\subsection{Function spaces}\label{sec:function_spaces}

\begin{table}[ht!]
\begin{center}
\begin{tabular}{cccc}
\multicolumn{4}{c}{Prognostic variables} \\ \hline
Field & Symbol & Equation & Function space \\ \hline
Water elevation & $\eta$ & \eqref{eq:freesurface_weak} & $\text{P}^{\text{DG}}_1$ \\
Depth av. velocity & $\bar{\mathbf{u}}$ & \eqref{eq:mom2d_weak} & $[\text{P}^{\text{DG}}_1]^2$ \\
Horizontal velocity & $\mathbf{u}'$ & \eqref{eq:mom_weak} & $[\text{P}^{\text{DG}}_1\times\text{P}^{\text{DG}}_1]^2$ \\
Water temperature & $T$ & \eqref{eq:tracer_weak} & $\text{P}^{\text{DG}}_1\times\text{P}^{\text{DG}}_1$ \\
Water salinity & $S$ & \eqref{eq:tracer_weak} & $\text{P}^{\text{DG}}_1\times\text{P}^{\text{DG}}_1$ \\
\multicolumn{4}{c}{Diagnostic variables} \\ \hline
Field & Symbol & Equation & Function space \\ \hline
Vertical velocity & $w$ & \eqref{eq:continuity_weak} & $\text{P}^{\text{DG}}_1\times\text{P}^{\text{DG}}_1$ \\
Water density & $\rho'$ & \eqref{eq:lin_eos} & $\text{P}^{\text{DG}}_1\times\text{P}^{\text{DG}}_1$ \\
Baroclinic head & $r$ & \eqref{eq:weak_r_eq} & $\text{P}^{\text{DG}}_1\times\text{P}_2$ \\
Int. pressure grad. & $\mathbf{F}_{\text{pg}}$ & \eqref{eq:weak_int_pg_eq} & $[\text{P}^{\text{DG}}_1\times\text{P}^{\text{DG}}_1]^2$ \\
\end{tabular}
\end{center}
\caption{
Prognostic and diagnostic variables and their function spaces.
}\label{tab:variables}
\end{table}

The prognostic variables of the coupled 2D--3D system (\ref{eq:mom2d},\ref{eq:freesurface},\ref{eq:mom3d},\ref{eq:tracer}) are $\eta,\bar{\mathbf{u}},\mathbf{u}',T$, and $S$.
Diagnostic variables include the vertical velocity $w$, water density $\rho'$, baroclinic head $r$, and internal pressure gradient $\mathbf{F}_{\text{pg}}$.
The choice of function spaces where these variables reside is crucial for numerical stability and accuracy.

Our discretization is based on the linear discontinuous Galerkin function space, $\text{P}^{\text{DG}}_1$.
The 2D system is discretized with a $\text{P}^{\text{DG}}_1-\text{P}^{\text{DG}}_1$ velocity--pressure finite element pair:
Water elevation and both components of the depth-averaged velocity are approximated in $\text{P}^{\text{DG}}_1$ space, i.e. $\eta\in\mathcal{H}_{\text{2D}}=\text{P}^{\text{DG}}_1$, $\mathbf{u}\in\mathcal{U}_{\text{2D}}=[\text{P}^{\text{DG}}_1]^2$.
When embedded with appropriate Riemann fluxes at element interfaces the $\text{P}^{\text{DG}}_1-\text{P}^{\text{DG}}_1$ element pair is well suited for rotational shallow water problems \citep{comblen2010,karna2011}.

Achieving an accurate and monotone 3D tracer advection scheme is one of our main design criteria.
The tracers therefore are also considered within a discontinuous function space, $T,S\in\mathcal{H}=\text{P}^{\text{DG}}_1\times\text{P}^{\text{DG}}_1$ (here the $\times$ operator stands for the Cartesian product of function spaces in the extruded mesh: horizontal $\times$ vertical function space).
Tracer consistency (sometimes called local tracer conservation) is a necessary condition for monotonicity;
it ensures that a constant tracer field does not exhibit spurious local extrema.
In practice it implies that the discrete tracer equation must reduce to the discrete continuity equation for a constant tracer.
In this work we satisfy this property by requiring that the vertical velocity belongs to the tracer space $\mathcal{H}$ \citep{white2008}.
In addition, compatibility between the 2D and 3D momentum equations requires that the 3D horizontal velocity must be $\text{P}^{\text{DG}}_1$ in the horizontal direction.
We therefore set $\mathbf{u}'\in\mathcal{U}=[\text{P}^{\text{DG}}_1\times\text{P}^{\text{DG}}_1]^2$ as well.

Note that this choice of function spaces is not \emph{mimetic} \citep{mcrae2014,danilov2013}:
the discrete system does not preserve all the properties of the continuous equations, for example enstrophy is not conserved exactly.
As the coastal ocean is generally very dissipative, maintaining mimetic properties is however not crucial.
It is possible to define a mimetic discretization as well, for example using Raviart-Thomas elements for the velocity, i.e. element pair $\text{RT}_1-\text{P}^{\text{DG}}_1$ \citep{mcrae2014}.
Our preliminary experiments however indicate that this choice significantly increases the computational cost of the system, without a corresponding improvement in accuracy.
Formal assessment of the performance of mimetic discretizations in
coastal ocean applications will be investigated in the future.

In the weak forms we use the following notation for volume and interface integrals

\begin{align}
 \Big\langle \bullet  \Big\rangle_{\!\! \Omega} &= \int_{\Omega} \bullet\; \mathrm{d}\mathbf{x}, \\
 \Big\langle\!\!\!\Big\langle \Big\rangle\!\!\!\Big\rangle_{\!\!\partial\Omega} &= \int_{\partial\Omega} \bullet\; \mathrm{d}s.
\end{align}

In interface terms we additionally use the average $\{\!\!\!\{\cdot\}\!\!\!\}$ and jump $[\![\cdot]\!]$ operators for scalar $a$ and vector $\mathbf{u}$ fields:

\begin{align}
 \{\!\!\!\{a\}\!\!\!\} &= \frac{1}{2}(a^+ + a^-), \\
 \{\!\!\!\{\mathbf{u}\}\!\!\!\} &= \frac{1}{2}(\mathbf{u}^+ + \mathbf{u}^-), \\
 [\![a \mathbf{n}]\!] &= a^+ \mathbf{n}^+ + a^- \mathbf{n}^-, \\
 [\![\mathbf{u} \cdot \mathbf{n}]\!] &= \mathbf{u}^+ \cdot \mathbf{n}^+ + \mathbf{u}^- \cdot \mathbf{n}^-, \\
 [\![\mathbf{u} \mathbf{n}]\!] &= \mathbf{u}^+ \mathbf{n}^+ + \mathbf{u}^- \mathbf{n}^-,
\end{align}
where the superscripts '$+$' and '$-$' arbitrarily label the values on
either side of the interface and $\mathbf{n}$ is the outward unit
normal vector of each element on the interface.

\subsection{2D system}\label{sec:weak_2d_system}

Let $\mathcal{T}$ stand for the triangulation of the 2D domain $\Gamma_0$.
The set of element interfaces are denoted by $\mathcal{I} = \{k \cap k' \vert k,k' \in \mathcal{T}\}$, and $\mathbf{n}=(n_x,n_y)$ the outward unit normal vector of an interface $e\in\mathcal{I}$. For brevity boundary conditions are omitted from the weak forms.

Let $\phi_{\text{2D}}\in\mathcal{H}_{\text{2D}}$ and $\boldsymbol{\psi}_{\text{2D}}\in\mathcal{U}_{\text{2D}}$ be test functions in the 2D function spaces.
The weak formulation of the 2D system then reads, find $\eta \in \mathcal{H}_{\text{2D}}, \bar{\mathbf{u}} \in \mathcal{U}_{\text{2D}}$ such that

\begin{align}
 \Big\langle \frac{\partial \eta}{\partial t}\phi_{\text{2D}} \Big\rangle_{\!\!\Gamma_0} +
  \Big\langle\!\!\!\Big\langle \left(H^*\bar{\mathbf{u}}^*\right)\cdot [\![\phi_{\text{2D}}\mathbf{n}]\!] \Big\rangle\!\!\!\Big\rangle_{\!\!\mathcal{I}}
  -
  \Big\langle \left(H\bar{\mathbf{u}}\right)\cdot\boldsymbol{\nabla}_h\phi_{\text{2D}} \Big\rangle_{\!\!\Gamma_0}
  &= 0, \label{eq:freesurface_weak} \\
 \Big\langle \frac{\partial \bar{\mathbf{u}}}{\partial t}\cdot\boldsymbol{\psi}_{\text{2D}} \Big\rangle_{\!\!\Gamma_0}
 +
  \Big\langle f\mathbf{e}_z\wedge\bar{\mathbf{u}}\cdot\boldsymbol{\psi}_{\text{2D}} \Big\rangle_{\!\!\Gamma_0}
  +
  \Big\langle\!\!\!\Big\langle g \eta^* [\![\boldsymbol{\psi}_{\text{2D}} \cdot \mathbf{n}]\!] \Big\rangle\!\!\!\Big\rangle_{\!\!\mathcal{I}}
  -
  \Big\langle g\eta\boldsymbol{\nabla}_h\cdot\boldsymbol{\psi}_{\text{2D}} \Big\rangle_{\!\!\Gamma_0}
  &=
  \Big\langle \mathbf{G}\cdot\boldsymbol{\psi}_{\text{2D}} \Big\rangle_{\!\!\Gamma_0}
  ,\ \forall \phi_{\text{2D}} \in \mathcal{H}_{\text{2D}}, \boldsymbol{\psi}_{\text{2D}} \in \mathcal{U}_{\text{2D}}. \label{eq:mom2d_weak}
\end{align}

Here the divergence $\boldsymbol{\nabla}_h\cdot\left(H\bar{\mathbf{u}}\right)$ and external gradient $g\boldsymbol{\nabla}_h \eta$ terms have been integrated by parts.
The resulting interface terms are defined on the element edges where the state variables $\eta,\bar{\mathbf{u}}$ are not uniquely defined.
The values $\eta^*,\bar{\mathbf{u}}^*$ are obtained from an approximate Riemann solver;
here we use the linear Roe solution $\eta^* = \{\!\!\!\{\eta\}\!\!\!\} + \sqrt{H/g} [\![\bar{\mathbf{u}}\cdot\mathbf{n}]\!]$ and $\bar{\mathbf{u}}^* =  \{\!\!\!\{\bar{\mathbf{u}}\}\!\!\!\} + \sqrt{g/H} [\![\eta\mathbf{n}]\!]$ \citep{comblen2010}.

\subsection{Momentum equation}\label{sec:weak_mom_eq}

Let $\mathcal{P}$ denote the set of prisms of the 3D domain $\Omega$, obtained from a vertical extrusion of $\Gamma_0$.
The set of horizontal and vertical interfaces are denoted by $\mathcal{I}_h$ and $\mathcal{I}_v$, respectively.
Let $\boldsymbol{\psi} \in \mathcal{U}$ be a test function.
The weak formulation of the 3D momentum equation then reads: find $\mathbf{u} \in \mathcal{U}$ such that

\begin{align}
 \begin{split}
 \Big\langle \frac{\partial \mathbf{u}'}{\partial t}\cdot\boldsymbol{\psi} \Big\rangle_{\!\!\Omega}
   &-
   \Big\langle \boldsymbol{\nabla}_h \boldsymbol{\psi} : (\mathbf{u} \mathbf{u}) \Big\rangle_{\!\!\Omega}
   +
   \Big\langle\!\!\!\Big\langle \mathbf{u}^{\text{up}} \cdot [\![\boldsymbol{\psi} \mathbf{n}_h]\!] \cdot \!\!\{\mathbf{u}\}\!\!\!\} \Big\rangle\!\!\!\Big\rangle_{\!\!\mathcal{I}_{h}\cup\mathcal{I}_v}
   \\
   &-
   \Big\langle \left(w\mathbf{u} \right)\cdot\frac{\partial \boldsymbol{\psi}}{\partial z} \Big\rangle_{\!\!\Omega}
   +
   \Big\langle\!\!\!\Big\langle \mathbf{u}^{\text{up}} \cdot [\![\boldsymbol{\psi} n_z]\!] \{\!\!\!\{w\}\!\!\!\} \Big\rangle\!\!\!\Big\rangle_{\!\!\mathcal{I}_{h}}
   +
   \Big\langle f\mathbf{e}_z\wedge\mathbf{u}'\cdot\boldsymbol{\psi} \Big\rangle_{\!\!\Omega}
   \\
   &+
   \Big\langle \mathbf{F}_{\text{pg}} \cdot\boldsymbol{\psi} \Big\rangle_{\!\!\Omega}
   +
   \Big\langle\!\!\!\Big\langle \gamma_{\text{lf}} [\![\mathbf{u}]\!] \cdot [\![\boldsymbol{\psi}]\!] \Big\rangle\!\!\!\Big\rangle_{\!\!\mathcal{I}_{h}\cup\mathcal{I}_v}
 =
   D_h(\mathbf{u}, \boldsymbol{\psi})
   + D_v(\mathbf{u}, \boldsymbol{\psi}),\ \forall \boldsymbol{\psi} \in \mathcal{U}.
\end{split} \label{eq:mom_weak}
\end{align}

Here the advection and viscosity terms have been integrated by parts \citep[see][]{karna2013}; the colon operator is the Frobenius inner product, $\mathbf{A}:\mathbf{B}=\sum_{i,j}A_{i,j}B_{i,j}$, and $\mathbf{u}^{\text{up}}$ stands for the upwind value at the interface.
The internal pressure gradient term has been augmented with the Lax--Friedrichs flux with parameter $\gamma_{\text{lf}}=\{\!\!\!\{|\mathbf{u}|\}\!\!\!\}$.
Adding such a flux is required to stabilize the internal pressure gradient:
it reduces noise in the velocity field, and decreases spurious numerical mixing in baroclinic applications.
The $D_h,D_v$ terms denote the diffusion operators introduced later.

\subsection{Tracer equation}\label{sec:weak_tracer_eq}

The weak formulation of the tracer equations is derived analogously: find $T \in \mathcal{H}$ such that

\begin{align}
 \begin{split}
 \Big\langle \frac{\partial T}{\partial t}\phi \Big\rangle_{\!\!\Omega}
   &- \Big\langle T\mathbf{u} \cdot \boldsymbol{\nabla}_h\phi \Big\rangle_{\!\!\Omega}
   + \Big\langle\!\!\!\Big\langle T^{\text{up}} [\![\phi \mathbf{n}_h]\!] \cdot \{\!\!\!\{\mathbf{u}\}\!\!\!\} \Big\rangle\!\!\!\Big\rangle_{\!\!\mathcal{I}_h\cup\mathcal{I}_v} \\
   &- \Big\langle \left(Tw \right)\frac{\partial \phi}{\partial z} \Big\rangle_{\!\!\Omega}
   + \Big\langle\!\!\!\Big\langle T^{\text{up}} [\![\phi n_z]\!] \{\!\!\!\{w\}\!\!\!\} \Big\rangle\!\!\!\Big\rangle_{\!\!\mathcal{I}_v}
 =
  D_h(T, \phi)
  + D_v(T, \phi),\ \forall \phi \in \mathcal{H}.
   \label{eq:tracer_weak}
 \end{split}
\end{align}

Note that we do not employ the Lax--Friedrichs flux in the tracer equation. 

\subsection{Symmetric Interior Penalty stabilization}\label{sec:weak_sipg}

The presented discretization is unstable for elliptic operators, and the diffusion operators require additional stabilization.
Here we use the Symmetric Interior Penalty Galerkin (SIPG) method \citep{epshteyn2007}.
The SIPG formulation of the tracer diffusion operators read

\begin{align}
 &\begin{aligned}
 D_h(T, \phi) =
    &-
    \Big\langle \mu_h (\boldsymbol{\nabla}_h \phi) \cdot (\boldsymbol{\nabla}_h T) \Big\rangle_{\!\!\Omega}
    + \Big\langle\!\!\!\Big\langle \{\!\!\!\{\mu_h \boldsymbol{\nabla}_h T\}\!\!\!\} \cdot [\![\phi \mathbf{n}_h]\!] \Big\rangle\!\!\!\Big\rangle_{\!\!\mathcal{I}_h\cup\mathcal{I}_v} \\
    &+ \Big\langle\!\!\!\Big\langle \{\!\!\!\{\mu_h \boldsymbol{\nabla}_h \phi\}\!\!\!\} \cdot [\![T \mathbf{n}_h]\!] \Big\rangle\!\!\!\Big\rangle_{\!\!\mathcal{I}_h\cup\mathcal{I}_v}
    - \Big\langle\!\!\!\Big\langle \{\!\!\!\{\sigma\}\!\!\!\} \{\!\!\!\{\mu_h\}\!\!\!\} [\![T \mathbf{n}_h]\!] \cdot [\![\phi \mathbf{n}_h]\!] \Big\rangle\!\!\!\Big\rangle_{\!\!\mathcal{I}_h\cup\mathcal{I}_v},
 \end{aligned} \label{eq:weak_h_diff_term} \\
 &\begin{aligned}
 D_v(T, \phi) =
    &-
    \Big\langle \mu \frac{\partial T}{\partial z} \frac{\partial \phi}{\partial z} \Big\rangle_{\!\!\Omega}
    + \Big\langle\!\!\!\Big\langle \Big\{\!\!\!\!\Big\{\mu \frac{\partial T}{\partial z}\Big\}\!\!\!\!\Big\} [\![\phi n_z]\!] \Big\rangle\!\!\!\Big\rangle_{\!\!\mathcal{I}_{h}}
    \\
    &+ \Big\langle\!\!\!\Big\langle \Big\{\!\!\!\!\Big\{\mu \frac{\partial \phi}{\partial z}\Big\}\!\!\!\!\Big\} [\![T n_z]\!] \Big\rangle\!\!\!\Big\rangle_{\!\!\mathcal{I}_{h}}
    - \Big\langle\!\!\!\Big\langle \{\!\!\!\{\sigma\}\!\!\!\}\{\!\!\!\{\mu\}\!\!\!\} [\![T n_z]\!] [\![\phi n_z]\!] \Big\rangle\!\!\!\Big\rangle_{\!\!\mathcal{I}_{h}}.
 \end{aligned} \label{eq:weak_v_diff_term}
\end{align}
For the viscosity terms we get

\begin{align}
 &\begin{aligned}
 D_h(\mathbf{u}, \boldsymbol{\psi}) =
   &- \Big\langle \nu_h (\boldsymbol{\nabla}_h \boldsymbol{\psi}) : (\boldsymbol{\nabla}_h \mathbf{u})^T \Big\rangle_{\!\!\Omega}
   + \Big\langle\!\!\!\Big\langle [\![\boldsymbol{\psi} \mathbf{n}_h]\!] \cdot \{\!\!\!\{\nu_h \boldsymbol{\nabla}_h \mathbf{u}\}\!\!\!\} \Big\rangle\!\!\!\Big\rangle_{\!\!\mathcal{I}_h\cup\mathcal{I}_v} \\
   &+ \Big\langle\!\!\!\Big\langle [\![\mathbf{u} \mathbf{n}_h]\!] \cdot \{\!\!\!\{\nu_h \boldsymbol{\nabla}_h \boldsymbol{\psi}\}\!\!\!\} \Big\rangle\!\!\!\Big\rangle_{\!\!\mathcal{I}_h\cup\mathcal{I}_v}
   - \Big\langle\!\!\!\Big\langle \{\!\!\!\{\sigma\}\!\!\!\} \{\!\!\!\{\nu_h\}\!\!\!\} [\![\mathbf{u} \mathbf{n}_h]\!] [\![\boldsymbol{\psi} \mathbf{n}_h]\!] \Big\rangle\!\!\!\Big\rangle_{\!\!\mathcal{I}_h\cup\mathcal{I}_v},  \label{eq:weak_h_visc_term} \\
 \end{aligned} \\
 &\begin{aligned}
 D_v(\mathbf{u}, \boldsymbol{\psi}) =
   &- \Big\langle \nu \frac{\partial \boldsymbol{\psi}}{\partial z} \cdot \frac{\partial \mathbf{u}}{\partial z} \Big\rangle_{\!\!\Omega}
   + \Big\langle\!\!\!\Big\langle [\![\boldsymbol{\psi} n_z]\!] \cdot \Big\{\!\!\!\!\Big\{\nu \frac{\partial \mathbf{u}}{\partial z}\Big\}\!\!\!\!\Big\} \Big\rangle\!\!\!\Big\rangle_{\!\!\mathcal{I}_{h}} \\
   &+ \Big\langle\!\!\!\Big\langle [\![\mathbf{u} n_z]\!] \cdot \Big\{\!\!\!\!\Big\{\nu \frac{\partial \boldsymbol{\psi}}{\partial z}\Big\}\!\!\!\!\Big\} \Big\rangle\!\!\!\Big\rangle_{\!\!\mathcal{I}_{h}}
   - \Big\langle\!\!\!\Big\langle \{\!\!\!\{\sigma\}\!\!\!\} \{\!\!\!\{\nu\}\!\!\!\} [\![\mathbf{u} n_z]\!] \cdot [\![\boldsymbol{\psi} n_z]\!] \Big\rangle\!\!\!\Big\rangle_{\!\!\mathcal{I}_{h}}.  \label{eq:weak_v_visc_term}
 \end{aligned}
\end{align}
The penalty factor $\sigma$ is defined as $\sigma = \gamma \frac{p (p+1)}{L}$ \citep{epshteyn2007},
where $p$ is the degree of the basis functions, $\gamma$ is a factor depending on mesh quality, and $L$ is the local element length scale in the normal direction of the interface.
Let $h_h$ and $h_v$ denote the horizontal and vertical element sizes, and $\boldsymbol{\Delta}=\text{diag}(h_h,h_h,h_v)$.
We then define $L=\mathbf{n}\cdot\boldsymbol{\Delta}\cdot\mathbf{n}=h_h (n_x^2+n_y^2)+h_v n_z^2$ \citep{pestiaux2014}.
In this paper we use $\gamma=5$.

\subsection{Continuity equation}\label{sec:weak_w_eq}

The vertical velocity $w$ is computed diagnostically from the continuity equation \eqref{eq:continuity3d} by solving the weak form: find $w \in \mathcal{H}$ such that

\begin{align}
    \Big\langle\!\!\!\Big\langle w n_z \varphi \Big\rangle\!\!\!\Big\rangle_{\!\!\Gamma_s}
    + \Big\langle\!\!\!\Big\langle \{\!\!\!\{w\}\!\!\!\} [\![\varphi n_z]\!] \Big\rangle\!\!\!\Big\rangle_{\!\!\mathcal{I}_h}
    - \Big\langle w \frac{\partial \varphi}{\partial z} \Big\rangle_{\!\!\Omega}
    = \Big\langle \mathbf{u} \cdot \boldsymbol{\nabla}_h \varphi \Big\rangle_{\!\!\Omega}
    - \Big\langle\!\!\!\Big\langle \{\!\!\!\{\mathbf{u}\}\!\!\!\} \cdot [\![\varphi \mathbf{n}_h]\!] \Big\rangle\!\!\!\Big\rangle_{\!\!\mathcal{I}_h \cup \mathcal{I}_v}
    - \Big\langle\!\!\!\Big\langle \mathbf{u} \cdot \varphi \mathbf{n}_h \Big\rangle\!\!\!\Big\rangle_{\!\!\Gamma_s},\ \forall \varphi \in \mathcal{H}, \label{eq:continuity_weak}
\end{align}
where both the left and right hand sides have been integrated by parts.
Note that the terms on the bottom surface $\Gamma_b$ vanish due to the impermeability constraint $\mathbf{u}\cdot\mathbf{n}_h + wn_z = 0$.

\subsection{Computing the internal pressure gradient}\label{sec:weak_int_pg}

The water density is computed diagnostically using the equation of state.
We use the same $\text{P}^{\text{DG}}_1\times\text{P}^{\text{DG}}_1$ function space for tracers and water density.
In this work we use a linear equation of state \eqref{eq:lin_eos}, and consequently density can  be computed locally (at each node of the tracer field).
In general, however, the equation of state is non-linear, and the density is projected on the $\rho$ field.

The baroclinic head is computed from \eqref{eq:baroc_head} by integrating $\rho'$ over the vertical.
In practice we solve equation $\frac{\partial r}{\partial z} = \rho'/\rho_0$ weakly with the appropriate boundary conditions:
\begin{align}
    \Big\langle\!\!\!\Big\langle r n_z \varphi \Big\rangle\!\!\!\Big\rangle_{\!\!\Gamma_b}
    + \Big\langle\!\!\!\Big\langle r_\text{up} [\![\varphi n_z]\!] \Big\rangle\!\!\!\Big\rangle_{\!\!\mathcal{I}_h}
    - \Big\langle r \frac{\partial \varphi}{\partial z} \Big\rangle_{\!\!\Omega}
    = \Big\langle \frac{1}{\rho_0}\rho' \varphi \Big\rangle_{\!\!\Omega}. \label{eq:weak_r_eq}
\end{align}
Here the left hand side has been integrated by parts, and $r_\text{up}$ denotes the value on the prism above the interface.
Note that the free surface terms vanish because $r = 0$ on $\Gamma_s$ by definition.
We use function space $\text{P}^{\text{DG}}_1\times\text{P}_2$ for $r$ to alleviate internal pressure gradient errors \citep{piggott2008}.

Finally, taking a test function $\boldsymbol{\psi} \in \mathcal{U}$, we compute the internal pressure gradient with the weak form
\begin{align}
    \Big\langle \mathbf{F}_{\text{pg}} \cdot \boldsymbol{\psi} \Big\rangle_{\!\!\Omega}
    = - \Big\langle g r \boldsymbol{\nabla}_h \cdot \boldsymbol{\psi} \Big\rangle_{\!\!\Omega}
    + \Big\langle\!\!\!\Big\langle g \{\!\!\!\{r\}\!\!\!\} [\![\boldsymbol{\psi} \cdot \mathbf{n}_h]\!] \Big\rangle\!\!\!\Big\rangle_{\!\!\mathcal{I}_h \cup \mathcal{I}_v}
    + \Big\langle\!\!\!\Big\langle g r \boldsymbol{\psi} \cdot \mathbf{n}_h \Big\rangle\!\!\!\Big\rangle_{\!\!\Gamma_s \cup \Gamma_b},
    \ \forall \boldsymbol{\psi} \in \mathcal{U} \label{eq:weak_int_pg_eq}
\end{align}
where the right hand side has been integrated by parts.
Usually $\mathbf{F}_{\text{pg}}$ belongs to the same space as the horizontal velocity, i.e. $[\text{P}^{\text{DG}}_1\times\text{P}^{\text{DG}}_1]^2$.
However, to reduce bathymetry induced internal pressure gradient errors it is possible to use a quadratic horizontal space, i.e.  $r\in \text{P}^{\text{DG}}_2\times\text{P}_2$ and $\mathbf{F}_{\text{pg}} \in [\text{P}^{\text{DG}}_2\times\text{P}^{\text{DG}}_1]^2$.
In this paper we use a linear $\mathbf{F}_{\text{pg}}$ field unless otherwise specified.

\subsection{Slope limiters}\label{sec:slope_limiters}

We use vertex-based $\text{P}^{\text{DG}}_1$ slope limiters \citep{kuzmin2010} for three-dimensional variables to ensure positivity.
The limiter is applied to both tracer and horizontal velocity fields after each update of the advection operator as discussed in the next Section.

\section{Time integration}\label{sec:timeintegration}

The coupled 2D--3D system is advanced in time with a two-stage arbitrary Lagrangian Eulerian (ALE) time integration scheme.
In this section we present the ALE formulation and summarize the final time integration scheme.

\subsection{ALE mesh formulation}\label{sec:ale_formulation}

To accurately account for the free surface movement one must move the mesh in the vertical direction.
In this work we adopt the ALE method \citep{donea2004}.
Here we describe a mesh update procedure that stretches (or compresses) the mesh uniformly over the vertical direction.
The ALE formulation, however, allows more complex mesh moving methods as well, such as the (approximate) tracking of isopycnals \citep{hofmeister2010}.

In three dimensions an ALE update consists of solving an advection-diffusion equation between two domains, $\Omega^{n}$ and $\Omega^{n+1}$.
Here the domain is uniquely defined by the surface elevation field, such that for any time level $n$ the surface $\Gamma_{s}^{n}$ matches $\eta^{n}$.
Due to the chosen discretization the elevation field $\eta$ is discontinuous, yet we wish to maintain a conforming mesh, i.e. a continuous coordinate field $z$.
This is achieved by projecting the elevation field $\eta^n$ to a continuous space and updating the geometry with the continuous field $\eta_{\text{cg}}^n$.
The projection induces a small discrepancy between the elevation field and the 3D domain, but its effect remains negligible in practical applications because jumps in the elevation field are typically small.

Let $\Omega^{\text{ref}}$ be the reference domain corresponding to unperturbed elevation field $\eta_{\text{cg}} = 0$, and $z_{\text{ref}} \in [-h, 0]$ its vertical coordinate.
Applying a uniform mesh stretching, the time dependent mesh coordinates can then be written as

\begin{align}
 z^n = z_{\text{ref}} + \eta_{\text{cg}}^n \frac{z_{\text{ref}}+h}{h} \in [-h, \eta_{\text{cg}}^n]. \label{eq:mesh_z_coord}
\end{align}

The mesh velocity is obtained as $w_m = \frac{\partial z}{\partial t}$. In practice the consecutive fields $\eta_{\text{cg}}^{n+1}$ and $\eta_{\text{cg}}^{n}$ are known so we can evaluate

\begin{align}
 w_m^{n+1} = \frac{\eta_{\text{cg}}^{n+1}-\eta_{\text{cg}}^{n}}{\Delta t} \frac{z_{\text{ref}}+h}{h}. \label{eq:mesh_velocity}
\end{align}

Given the mesh velocity a conservative ALE update can be written as

\begin{align}
 \frac{d}{d t}\left( \Big\langle T\phi \Big\rangle_{\!\!\Omega} \right) &= \Big\langle F_T(T, \mathbf{u}, w-w_m)\phi \Big\rangle_{\!\!\Omega},
\end{align}
for a generic tracer equation $\frac{\partial T}{\partial t} = F_T(T, \mathbf{u}, w)$.

\subsection{Coupled time integration scheme}\label{sec:coupled_time_integrator}

The coupled 2D--3D system is advanced in time with a two-stage ALE time integration scheme.
For convenience we re-write the 3D momentum and tracer equations as

\begin{align}
 \frac{\partial T}{\partial t} &= F_T(T, \mathbf{u}, w) + G_T(T), \\
 \frac{\partial \mathbf{u}}{\partial t} &= F_{\mathbf{u}}(\mathbf{F}_{\text{pg}}, \mathbf{u}, w) + G_{\mathbf{u}}(\mathbf{u}),
\end{align}
where $F_T$ and $F_{\mathbf{u}}$ denote all the terms that are treated explicitly while $G_T$ and $G_{\mathbf{u}}$ contain all the implicit terms.
In this work only vertical diffusion \eqref{eq:weak_v_diff_term}, vertical viscosity \eqref{eq:weak_v_visc_term}, and bottom friction terms are treated implicitly.

The explicit 3D equations are advanced in time with a second-order strong stability preserving (SSP) Runge-Kutta scheme,  SSPRK(2,2) \citep{shu1988,gottlieb1998}.
For a generic problem $\frac{\partial c}{\partial t} = F(c)$ the scheme reads:

\begin{align}
 c^{(1)} &= c^n + \Delta t F(c^n),\\
 c^{n+1} &= c^n + \frac{1}{2}\Delta t F(c^n) + \frac{1}{2}\Delta t F(c^{(1)}).
\end{align}

When applied to the explicit 3D momentum and tracer equations,  \eqref{eq:mom_weak} and \eqref{eq:tracer_weak},
both of these stages are ALE updates where the mesh is updated from geometry $\Omega^n$ to $\Omega^{(1)}$ and then $\Omega^{n+1}$.
The ALE formulation of the explicit 3D tracer equation can then be written as
\begin{align}
 \Big\langle T^{(1)} \phi \Big\rangle_{\!\!\Omega^{(1)}}
 &=
 \Big\langle T^{n} \phi \Big\rangle_{\!\!\Omega^{n}}
 + \Delta t
 \Big\langle F_T(T^{n}, \mathbf{u}^{n}, w^{n}-w_m^{(1)})\phi \Big\rangle_{\!\!\Omega^{n}}
 \label{eq:tracer_weak_ale_stage1}, \\
 \begin{split}
 \Big\langle \widetilde{T}^{n+1} \phi \Big\rangle_{\!\!\Omega^{n+1}}
 &=
 \Big\langle T^{n} \phi \Big\rangle_{\!\!\Omega^{n}}
 + \frac{1}{2}\Delta t
 \Big\langle F_T(T^{n}, \mathbf{u}^{n}, w^{n}-w_m^{(1)})\phi \Big\rangle_{\!\!\Omega^{n}}
 + \\
 &\frac{1}{2}\Delta t
 \Big\langle F_T(T^{(1)}, \mathbf{u}^{(1)}, w^{(1)}-w_m^{n+1})\phi \Big\rangle_{\!\!\Omega^{{(1)}}},
 \end{split}\label{eq:tracer_weak_ale_stage2}
\end{align}
where the vertical velocity is adjusted by the mesh velocity $w_m$.

After the SSPRK update, the implicit terms are advanced with the backward Euler method. This step is computed in domain $\Omega^{n+1}$:
\begin{align}
 \Big\langle T^{n+1} \phi \Big\rangle_{\!\!\Omega^{n+1}}
 &=
 \Big\langle \widetilde{T}^{n+1} \phi \Big\rangle_{\!\!\Omega^{n+1}}
 + \Delta t
 \Big\langle G_T(T^{n+1})\phi \Big\rangle_{\!\!\Omega^{n+1}}
 . \label{eq:tracer_weak_impl_stage}
\end{align}
The 3D momentum equation is treated analogously.

\begin{algorithm}[th]
 \caption{Summary of the coupled time integration algorithm.}\label{alg:time_integration}
 \begin{algorithmic}[1]
    \REQUIRE Model state variables at time $t_n$: $\eta^{n},\bar{\mathbf{u}}^{n},T^{n},S^{n},\mathbf{u}'^{n}$
    \\ \hspace*{-18pt} \textbf{First stage:} \\
    \STATE Solve 2D system for $(\eta^{(1)},\bar{\mathbf{u}}^{(1)})$ \eqref{eq:freesurface_weak_ale_stage1}--\eqref{eq:mom2d_weak_ale_stage1}
    \STATE Update mesh geometry to $\Omega^{(1)}$ and compute mesh velocity $w_m^{(1)}$ \eqref{eq:mesh_velocity}
    \STATE Update 3D equations with ALE step for $T^{(1)},S^{(1)},\mathbf{u}'^{(1)}$ \eqref{eq:tracer_weak_ale_stage1}
    \STATE Apply slope limiter to $T^{(1)},S^{(1)},\mathbf{u}'^{(1)}$
    \STATE Update the 2D coupling term $\mathbf{G}$ \eqref{eq:uv_coupling} and correct $\mathbf{u}'$ \eqref{eq:uv_correction}
    \STATE Update $w$ \eqref{eq:continuity_weak}, water density \eqref{eq:lin_eos}, and pressure gradient \eqref{eq:weak_int_pg_eq}
    \\ \hspace*{-18pt} \textbf{Second stage:}\\
    \STATE Solve 2D system for $(\eta^{n+1},\bar{\mathbf{u}}^{n+1})$ \eqref{eq:freesurface_weak_ale_stage2}--\eqref{eq:mom2d_weak_ale_stage2}
    \STATE Update mesh geometry to $\Omega^{n+1}$ and compute mesh velocity $w_m^{n+1}$ \eqref{eq:mesh_velocity}
    \STATE Update 3D equations with ALE step for $\widetilde{T}^{n+1},\widetilde{S}^{n+1},\widetilde{\mathbf{u}}'^{n+1}$ \eqref{eq:tracer_weak_ale_stage2}
    \STATE Apply slope limiter to $\widetilde{T}^{n+1},\widetilde{S}^{n+1},\widetilde{\mathbf{u}}'^{n+1}$
    \\ \hspace*{-18pt} \textbf{Final stage:}\\
    \STATE Update the 2D coupling term $\mathbf{G}$ \eqref{eq:uv_coupling} and correct $\mathbf{u}'$ \eqref{eq:uv_correction}
    \STATE Solve vertical viscosity and diffusion implicitly \eqref{eq:tracer_weak_impl_stage}
    \STATE Update $w$ \eqref{eq:continuity_weak}, water density \eqref{eq:lin_eos}, and pressure gradient \eqref{eq:weak_int_pg_eq}
    \STATE Update parametrizations (e.g. bottom friction and viscosity)
 \end{algorithmic}
\end{algorithm}

The 2D equations are advanced in time with an implicit scheme to circumvent the strict time step constraint imposed by surface gravity waves.
To ensure consistency between the movement of the 3D mesh and the 2D mode, the 2D time integration scheme must be compatible with the aforementioned SSPRK(2,2) method.
Here we use a combination of a forward Euler and trapezoidal steps:

\begin{align}
 c^{(1)} &= c^n + \Delta t F(c^{n}) \label{eq:timeint_2d_stage1}, \\
 c^{n+1} &= c^n + \frac{1}{2}\Delta t \left( F(c^{n}) + F(c^{n+1}) \right).  \label{eq:timeint_2d_stage2}
\end{align}

Denoting the tendencies of the 2D system \eqref{eq:freesurface_weak}-\eqref{eq:mom2d_weak} by $F_{\eta}$ and $F_{\bar{\mathbf{u}}}$, respectively, 
we can write the 2D solver as

\begin{align}
 \Big\langle \eta^{(1)} \phi_{\text{2D}} \Big\rangle_{\!\!\Gamma_0} =&
 \Big\langle \eta^{n} \phi_{\text{2D}} \Big\rangle_{\!\!\Gamma_0}
 + \Delta t \Big\langle F_{\eta}(\eta^{n}, \bar{\mathbf{u}}^{n}) \phi_{\text{2D}} \Big\rangle_{\!\!\Gamma_0},  \label{eq:freesurface_weak_ale_stage1}\\
 \Big\langle \bar{\mathbf{u}}^{(1)}\cdot\boldsymbol{\psi}_{\text{2D}} \Big\rangle_{\!\!\Gamma_0}
 =& \Big\langle \bar{\mathbf{u}}^{n}\cdot\boldsymbol{\psi}_{\text{2D}} \Big\rangle_{\!\!\Gamma_0}
 + \Delta t \Big\langle F_{\bar{\mathbf{u}}}(\eta^{n}, \bar{\mathbf{u}}^{n})\cdot\boldsymbol{\psi}_{\text{2D}} \Big\rangle_{\!\!\Gamma_0},  \label{eq:mom2d_weak_ale_stage1}\\
 \Big\langle \eta^{n+1} \phi_{\text{2D}} \Big\rangle_{\!\!\Gamma_0} =&
 \Big\langle \eta^{n} \phi_{\text{2D}} \Big\rangle_{\!\!\Gamma_0}
 + \frac{\Delta t}{2} \Big\langle \Big( F_{\eta}(\eta^{n}, \bar{\mathbf{u}}^{n}) + F_{\eta}(H^{n}, \bar{\mathbf{u}}^{n+1}) \Big)\phi_{\text{2D}} \Big\rangle_{\!\!\Gamma_0},  \label{eq:freesurface_weak_ale_stage2}\\
 \Big\langle \bar{\mathbf{u}}^{n+1}\cdot\boldsymbol{\psi}_{\text{2D}} \Big\rangle_{\!\!\Gamma_0} =&
 \Big\langle \bar{\mathbf{u}}^{n}\cdot\boldsymbol{\psi}_{\text{2D}} \Big\rangle_{\!\!\Gamma_0}
 + \frac{\Delta t}{2} \Big\langle \Big( F_{\bar{\mathbf{u}}}(\eta^{n}, \bar{\mathbf{u}}^{n}) + F_{\bar{\mathbf{u}}}(\eta^{n+1}, \bar{\mathbf{u}}^{n+1}) \Big)\cdot\boldsymbol{\psi}_{\text{2D}} \Big\rangle_{\!\!\Gamma_0}. \label{eq:mom2d_weak_ale_stage2}
\end{align}
The second implicit stage is linearized by treating the total depth $H$ explicitly in \eqref{eq:freesurface_weak_ale_stage2}.

The 2D system is solved first, resulting in an updated elevation field ($\eta^{(1)}$ and $\eta^{n+1}$ for the two stages, respectively)
and consequently mesh geometry ($\Omega^{(1)}$ and $\Omega^{n+1}$).
Once the mesh geometry is known it is straightforward to compute the corresponding mesh velocity $w_m$ and perform a 3D ALE update.

The time integration method is second-order for all the terms.
The whole algorithm is summarized in Algorithm \ref{alg:time_integration}.

\subsection{Choosing the time step}

The maximal admissible time step is constrained by the stability of the coupled time integrator.
The presented SSPRK(2,2) scheme has a CFL (Courant--Friedrichs--Lewy) factor 1.
The 2D scheme \eqref{eq:timeint_2d_stage2}, and the implicit vertical solver \eqref{eq:tracer_weak_impl_stage}, on the other hand, are unconditionally stable.
This implies that the coupled system is stable under the same conditions as the explicit SSP scheme on its own. 

The horizontal advection term imposes a constraint

\begin{align}
 \Delta t_{\text{adv}} =  \frac{\sigma_h L_h}{U},
 \label{eq:dt_hor_adv}
\end{align}
where $L_h$ is the horizontal element size, $U$ is the maximal horizontal velocity scale, and $\sigma_h$ is a length scaling factor.
For the presented $\text{P}^{\text{DG}}_1$ discretization we take $L_h$ as the square root of the triangle area.
For rectangular $\text{P}^{\text{DG}}_1$ elements and 2nd order RK schemes the scaling factor is approximately $\sigma_h = 0.33$ \citep{cockburn2001}.
In this work we use $\sigma_h = 0.05$ for all the diagnostic test cases.
In strongly stratified flows internal waves may impose a stricter constraint
\begin{align}
 \Delta t_{\text{iw}} = \frac{\sigma_h L_h}{C_{\text{iw}} + U},
 \label{eq:dt_int_wave}
\end{align}
where $C_{\text{iw}}$ is the speed of the internal waves.

Analogously, the time step constraint for vertical advection is
\begin{align}
 \Delta t_{\text{w}} =  \frac{\sigma_v L_z}{W},
 \label{eq:dt_vert_adv}
\end{align}
where $L_z$ is the element height, $W$ is the vertical velocity scale, and $\sigma_v = 0.125$ is the scaling factor. 

Given a horizontal viscosity scale $N_h$, the explicit viscosity operator imposes a constraint
\begin{align}
 \Delta t_{\text{visc}} = \sigma_{\text{visc}} \frac{(\sigma_h L_h)^2}{N_h}.
 \label{eq:dt_hor_visc}
\end{align}
which may become stringent for small elements and large viscosity values.
The scaling factor $\sigma_{\text{visc}}$ depends on the used stabilization scheme; here a value $\sigma_{\text{visc}} = 2$ is used.
The constraint for horizontal diffusivity is analogous.

In the simulations presented herein, the minimal admissible time step is evaluated on the mesh based on constant \emph{a-priori} velocity and viscosity scales.
The time step is kept constant throughout the simulation.

\section{Test cases}\label{sec:testcases}

We demonstrate the performance of the proposed discretization with a suite of test cases of increasing complexity.
We first evaluate the conservation and convergence of the solver in a barotropic standing wave test case.
The convergence of baroclinic terms is then examined in a specific steady-state test case.
The baroclinic solver and its numerical mixing is then evaluated with a non-rotating lock exchange test case and a rotating baroclinic eddy test, followed by the DOME overflow test.

\subsection{Standing wave}

We first evaluate the performance of the solver in a barotropic standing wave test case.
The domain is a $L_x=60$ \unit{km} long rectangular channel, 625 \unit{m} wide, and 100 \unit{m} deep. All lateral boundaries are closed.
Initially the water is at rest. A 10 \unit{m} tall sinusoidal elevation perturbation is prescribed along the channel ($\eta_a=-\eta_0\;\cos(2 \pi x/ L_x),$ $\eta_0=10$ \unit{m}), leading to a nonlinear wave as the simulation progresses.

The simulation is run for two full wave periods, approximately $3831.31$ \unit{s}.
To investigate tracer conservation and consistency properties two passive tracers are included:
salinity is set to a constant 4.5 \unit{psu}, while temperature varies between $5.0$ and $15.0$ \unit{^\circ C} along the channel ($T=5 \sin(2 \pi x/L_x) + 10$  \unit{^\circ C}).

The domain was discretized with a split-quad mesh using 40 elements along the channel (1500 \unit{m} edge length) and 4 vertical layers.
The time step is $\Delta t=95.78$ \unit{s}, chosen to meet the horizontal advection condition.

\begin{figure}[th]
\centering
\includegraphics[width=0.8\textwidth]{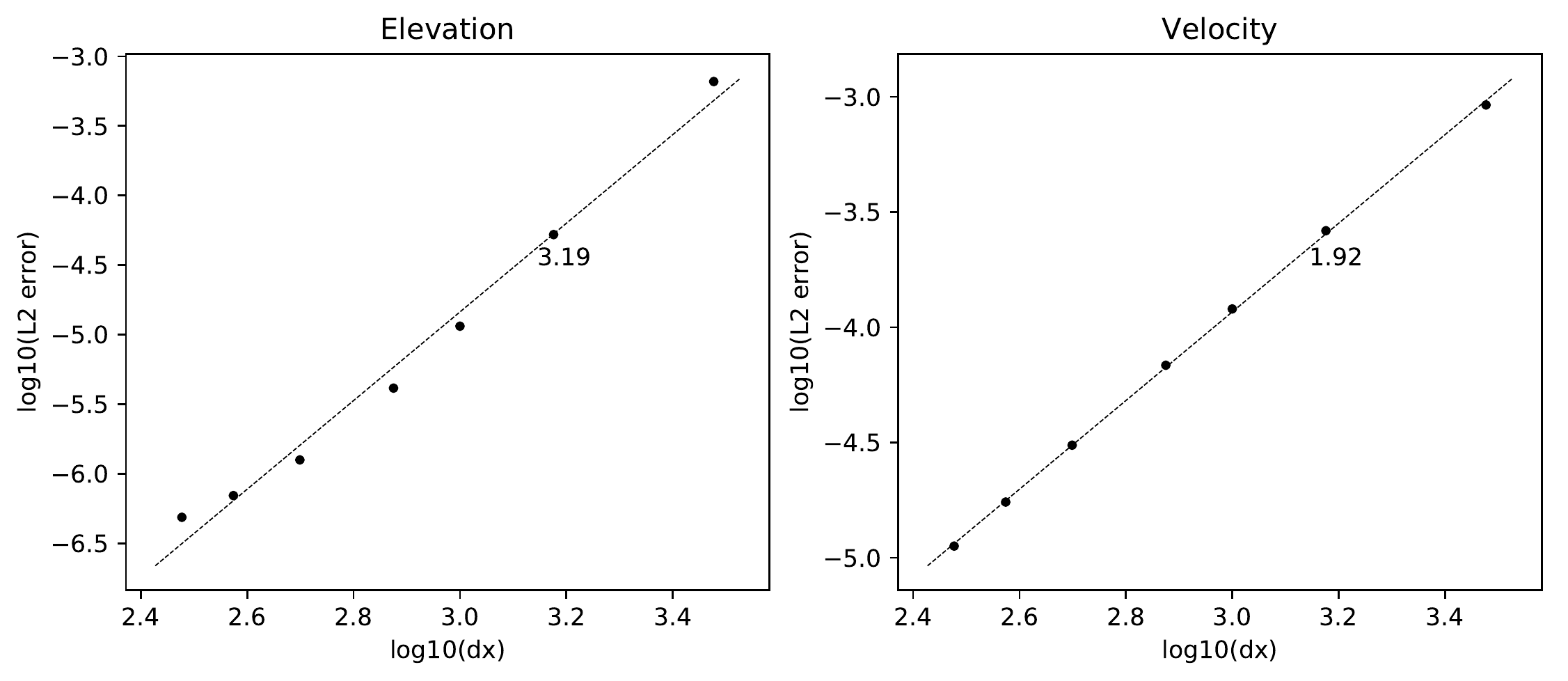}
\caption{
Convergence of the $L_2$ error in the standing wave test case.
Tested element sizes were 3000, 1500,  1000, 750, 500, 375, and  300 \unit{m}.
The number indicates the slope of the least-squares best fit line (dashed line).
}
\label{fig:standingwave_convergence}
\end{figure}

During the simulation the volume of the 3D domain was conserved to accuracy $\mathcal{O}(10^{-15})$.
The ``2D volume'', i.e. the integral of the elevation field, was conserved to accuracy $\mathcal{O}(10^{-16})$.
Salinity remained at constant 4.5 \unit{psu} with a small $\mathcal{O}(10^{-9})$ deviation.
The total mass of salinity and temperature were both conserved to accuracy $\mathcal{O}(10^{-12})$.
Over- and undershoots in the temperature field were negligible due to the slope limiters.
Without the limiter temperature overshoots were $\mathcal{O}(10^{-2})$.
These results show that the model indeed fully conserves volume and tracers and does not exhibit overshoots.
Moreover, the tracer consistency property is satisfied, verifying the integrity of the ALE formulation.

To investigate the order of convergence of the solver, we used a smaller initial elevation perturbation $\eta_0 = 1$ \unit{cm}.
In this case the resulting standing wave is close to linear.
At the end of the simulation the solution was compared to the analytical solution of the linear wave equation (which coincides with the initial condition) by computing the $L_2$ error, $\mathcal{E}(\eta) = (\int_\Omega (\eta - \eta_a)^2 \text{d}\mathbf{x})^{1/2}$.

We ran the simulation varying the horizontal mesh resolution between 3 \unit{km} and 300 \unit{m}; the number of vertical levels varied between 2 and 20.
In each case the channel was made one element wide, and the time step was chosen to meet the CFL criterion for horizontal advection.
At the end of the simulation the $L_2$ error was computed for water elevation and velocity (see Figure \ref{fig:standingwave_convergence}).
The velocity field shows the expected second-order convergence, whereas elevation converges at a rate of 3.2.
It is known that $\text{P}^{\text{DG}}_1$ shallow water equations models may exhibit superconvergence properties, especially for the elevation field \citep{bernard2008,comblen2010}.
Here our results verify that the solver behaves as expected and yields second-order accuracy under barotropic forcing.

\subsection{Baroclinic MMS test}\label{sec:baroclinic_mms}

Verifying model accuracy under baroclinic forcing is more challenging as no analytical solutions are available.
Here we use the method of manufactured solutions \citep[MMS;][]{salari2000} to construct a steady state test case that allows us to verify the correctness of the discrete baroclinic operators.
The domain is a $L_x=15$ by $L_y=10$ \unit{km} large and $h=40$ \unit{m} deep rectangular box.
All lateral boundaries are closed.
We prescribe initial velocity and temperature fields
\begin{align}
 u_a &= \frac{1}{2} \sin{\left (\frac{2 \pi}{L_{x}} x \right )} \cos{\left (\frac{3 z}{h} \right )}, \\
 v_a &= \frac{1}{3} \cos{\left (\frac{\pi y}{L_{y}} \right )} \sin{\left (\frac{z}{2 h} \right )}, \\
 T_a &= 15 \sin{\left (\frac{\pi x}{L_{x}} \right )} \sin{\left (\frac{\pi y}{L_{y}} \right )} \cos{\left (\frac{z}{h} \right )} + 15.
\end{align}

These functions were chosen to be analytic (infinitely differentiable) and fully three-dimensional to better quantify the spatial discretization error.

Salinity is set to a constant $35$ \unit{psu}.
We use the linear equation of state \eqref{eq:lin_eos} with $\rho_0 = 1000$ \unit{kg\ m^{-3}}, $\alpha_T=0.2$ \unit{kg\ m^{-3}\ {^{\circ}} C^{-1}} and $T_0 = 5$ \unit{^{\circ} C^{-1}}.
For the sake of simplicity, bathymetry is constant and elevation is set to zero initially.
Coriolis frequency was set to a constant $f=10^{-4}$ \unit{s^{-1}}.
Bottom friction, viscosity, and diffusivity are omitted.

Without any additional forcing the initial conditions lead to a time-dependent solution.
Following the MMS strategy, we add analytical source terms in the dynamic equations that cancel all the active terms in the equations, leading to a steady state solution.
The remaining error is purely the discretization error of the advection, pressure gradient, and Coriolis operators.
The source terms are derived analytically and projected to the corresponding function space. The analytical formulae are given in Appendix \ref{sec:baroclinic_mms_source_terms}.

\begin{figure}[th]
\centering
\includegraphics[width=0.8\textwidth]{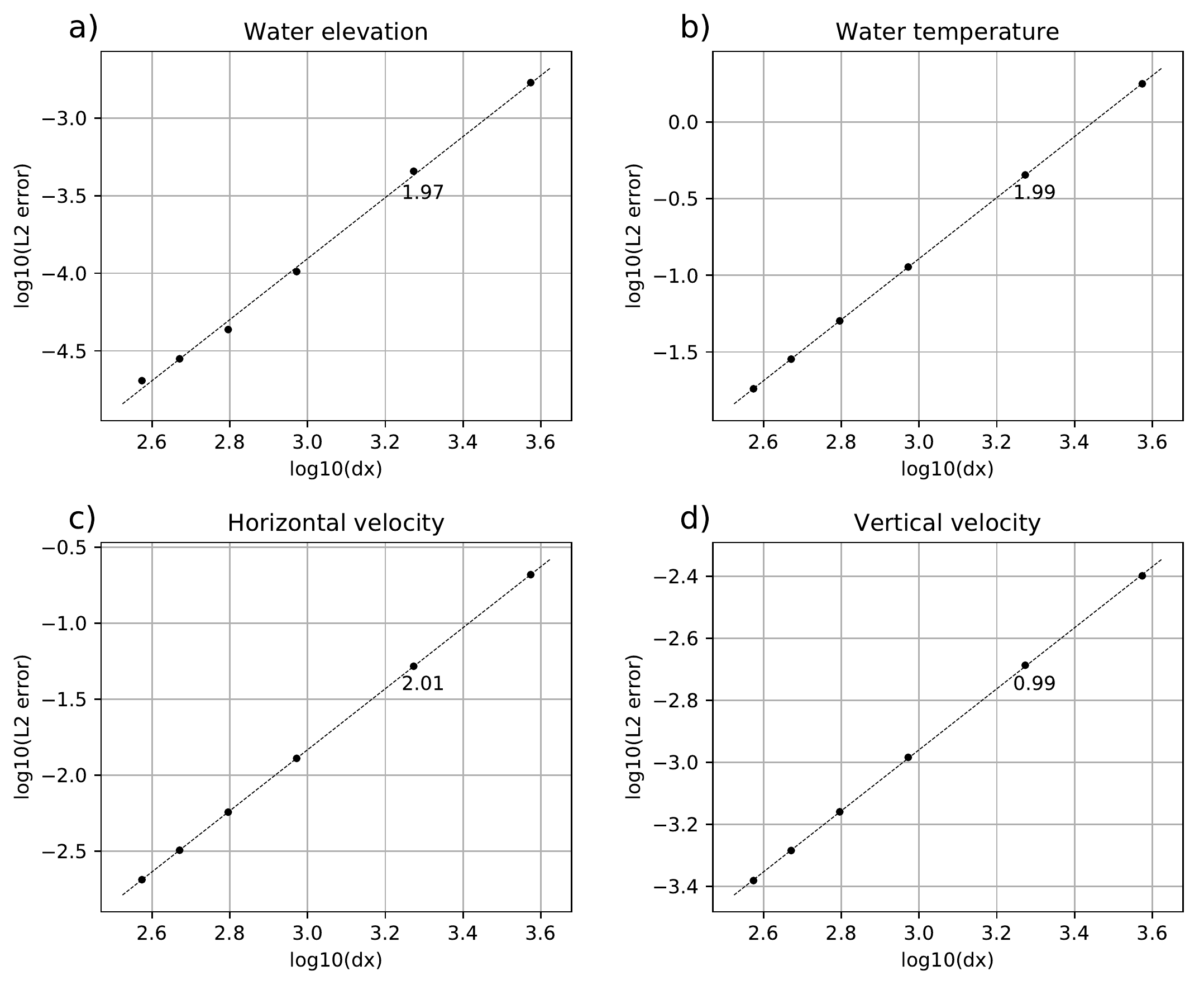}
\caption{
Convergence of the $L_2$ error in the baroclinic MMS test case.
The mesh was refined 1, 2, 4, 6, 8, and 10 times, resulting in resolution 2500, 1250, 625, 416.67, 312.5, and 250 \unit{m} (shortest edge of the triangle).
The time step was 25.0, 12.5, 6.25, 4.167, 3.125, and 2.5 \unit{s}, respectively.
The number indicates the slope of the least-squares best fit line (dashed line).
}
\label{fig:baroclinic-mms_convergence}
\end{figure}

The coarsest mesh contains 4 elements both in $x$ and $y$ directions and 2 vertical levels.
We refine the mesh up to 10 times (40 elements and 20 vertical levels) and compute the $L_2$ error of the prognostic fields against the exact solutions.
In each case, the model is run for 50 iterations with a time step chosen to meet the CFL condition.

The variation of the $L_2$ errors with resolution is shown in Figure \ref{fig:baroclinic-mms_convergence}.
All the prognostic variables exhibit the correct second-order convergence rate.
The diagnostic vertical velocity field (which depends on the divergence of $\mathbf{u}$) converges linearly as expected.
Therefore we conclude that advection, pressure gradient, and Coriolis terms are discretized correctly.
We have also developed similar MMS tests for the diffusivity/viscosity operators and the bottom friction term, all of which show second-order convergence as well (not shown).

\subsection{Lock exchange}

\begin{figure}[th]
\centering
\includegraphics[width=0.6\textwidth]{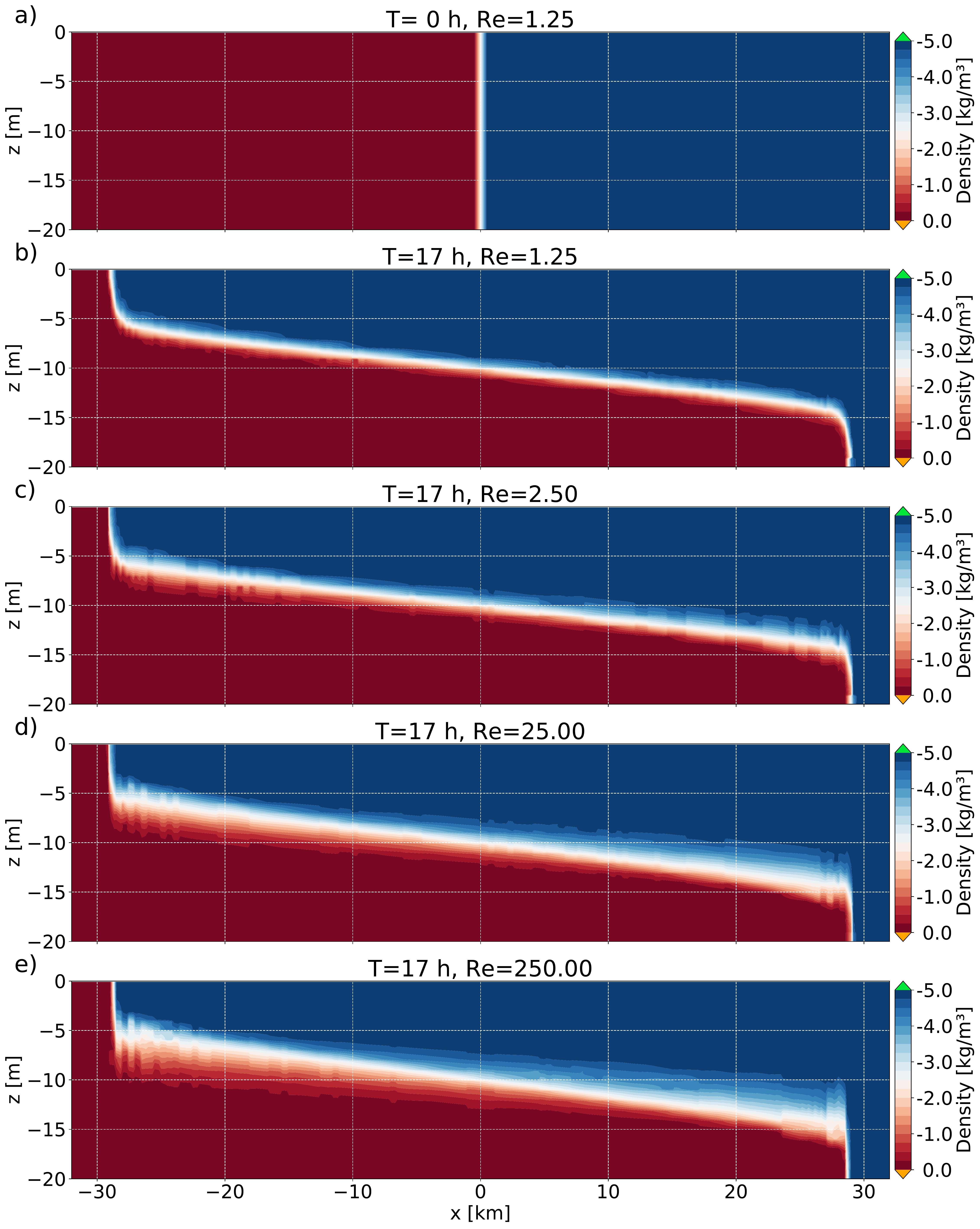}
\caption{
Water density in the lock exchange test case in the center of the domain ($y=0$ \unit{km}).
(a) Initial condition. Solution after 17 \unit{h} of simulation with $\text{Re}_h$ (b) 1.25, (c) 2.5, (d) 25.0, and (e) 250.0.
}
\label{fig:lock_test_solution}
\end{figure}

The validity of the baroclinic solver and its level of spurious mixing is investigated with the standard lock exchange test case \citep{wang1984,haidvogel1999,jankowski1999,ilicak2012,karna2013,petersen2015}.
Here we follow the setup of \cite{ilicak2012} and \cite{petersen2015}:
The domain is a 64 \unit{km} long and 1 \unit{km} wide rectangular channel. Water depth is 20 m.
Initially, the left-hand side of the domain is filled with dense water mass ($T=5.0$ \unit{^\circ C}) compared to the water on the right ($T=30.0$ \unit{^\circ C}).
Salinity is kept at constant $35$ \unit{psu}.
We use the same linear equation of state as in Section \ref{sec:baroclinic_mms},
resulting in a density difference of $\Delta \rho=5.0$ \unit{kg\ m^{-3}}.
The domain is discretized with a regular split-quad mesh.
The triangle edge length is 500 \unit{m} and 20 equidistant $\sigma$ levels are used in the vertical direction.

Stabilizing the internal pressure gradient requires some form of friction. 
To this end, we apply a constant Laplacian horizontal viscosity, using values $\nu = 1.0,\ 10.0,\ 100.0,\ \text{and}\ 200.0$ \unit{m^2\ s^{-1}}.
These values correspond to grid Reynolds numbers $\text{Re}_h = U \Delta x / \nu = 250.0,\ 25.0,\ 2.5,\ \text{and}\ 1.25$, respectively, where the characteristic velocity scale of the exchange flow is $U=0.5$ \unit{m\ s^{-1}}.
Vertical viscosity is set to a constant $10^{-4}$ \unit{m^2\ s^{-1}}.
Bottom friction is omitted.

Figure \ref{fig:lock_test_solution} shows the initial density field and solution after 17 h of simulation for the three cases.
Higher background viscosity leads to a less noisy velocity field and therefore sharper density front.
The sharpness and shape of the fronts are similar to results presented in the literature \citep[e.g. Fig. 5 in][]{ilicak2012}.
The low viscosity cases ($\text{Re}_h=25,250$) exhibit an internal wave at the front which significantly increases the overall mixing.

\begin{figure}[th]
\centering
\includegraphics[width=0.9\textwidth]{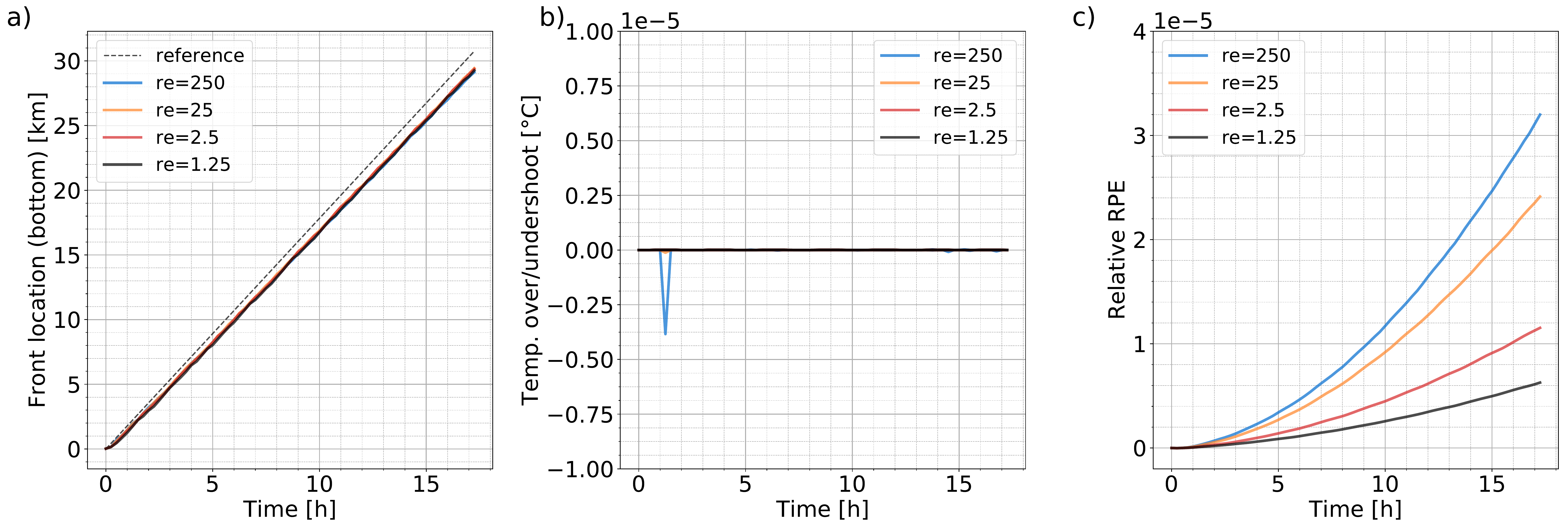}
\caption{
Diagnostics of the lock exchange test.
(a) Location of the density front at the bottom of the domain,
(b) Over- and undershoots in the temperature field (wrt. to $30.0$ and $5.0$ \unit{^\circ C}, respectively),
(c) Normalized reference potential energy (RPE) versus simulation time.
}
\label{fig:lock_test_diagnostics}
\end{figure}

Assuming that, in the absence of bottom friction, all available potential energy is transformed into kinetic energy, the maximum front propagation speed can be estimated as
$c=1/2\sqrt{g H\Delta\rho/\rho_0}$ \citep{benjamin1968,jankowski1999}.
Figure \ref{fig:lock_test_diagnostics} (a) shows the propagation of the front location at the bottom of the domain (the front at the surface behaves comparably).
All three simulations are in good agreement with the theoretical propagation speed.
The simulated front propagation speed is underestimated by roughly 5\% indicating loss of energy due to mixing.
These results are similar to results reported in the literature;
e.g. \cite{ilicak2012} show similar performance for ROMS, MITgcm, and MOM.

Figure \ref{fig:lock_test_diagnostics} (b) shows the maximum over- and undershoots in the temperature field during the simulation.
Even in the low viscosity case ($\text{Re}_h=250$), the overshoots are of order $10^{-5}$ \unit{^\circ C} indicating that the tracer advection scheme is indeed close to monotone, due to the SSP time integration method and slope limiters.
Note that if the slope limiter is omitted, the overshoots can reach 30 \unit{^\circ C}.

To diagnose the role of spurious mixing we use the reference potential energy (RPE, \citealt{ilicak2012,petersen2015}).
RPE is computed as the vertical center of mass of a sorted density field $\rho^*$: $RPE = g\int \rho^* (z+h) d\mathbf{x}$.
The $\rho^*$ field is defined as the unique, stratified density field where the densest water parcels are distributed over the bottom, and density increases monotonically over the water column.
As such, $\rho^*$ is the steady-state density distribution, and RPE represents the portion of potential energy that cannot be transformed into kinetic energy.
Mixing the two water masses increases RPE (the center of mass) and thus the amount of unavailable potential energy increases.
Figure \ref{fig:lock_test_diagnostics} (c) shows the evolution of normalized RPE, $\overline{RPE}(t) = (RPE(t)-RPE(0))/RPE(0)$ during the simulation.
At $t=17\ \text{h}$ the values are $0.612,\ 1.13,\ 2.35,\ 3.11\ \times10^{-5}$ for the four simulations.
These results are in good agreement with those reported with MPAS-Ocean model \citep{petersen2015}:
With the same mesh resolution, MPAS-Ocean shows slightly larger normalized RPE, for example, at $t=17\ \text{h}$ $\overline{RPE}\approx3.5\ \times10^{-5}$ in the case of $\text{Re}_h=25$.
The difference is likely due to the different spatial discretization ($\text{P}^{\text{DG}}_1$ instead of finite volumes), or differences in the numerical viscosity operator.
Applying slope limiters to the velocity field is not necessary for numerical stability, but it reduces high-frequency noise in the velocity field and hence results in lower RPE values.

In order to investigate the role of the Lax--Friedrichs flux on numerical mixing we ran the lock exchange test case with zero viscosity.
After 17 h of simulation, the RPE value was approximately $3.2\ \times10^{-5}$.
When the Lax--Friedrichs flux was omitted, a similar RPE value was obtained with viscosity $\nu = 3.125$ \unit{m^2\ s^{-1}}.
Therefore, in this particular test case, the Lax--Friedrichs flux introduces mixing that is roughly equivalent to $3$ \unit{m^2\ s^{-1}} viscosity, corresponding to $\text{Re}_h = 80$.
When viscosity was non-zero, it was evident from the numerical simulations that the Lax--Friedrichs flux has a negligible impact on numerical mixing if $Re<10$ (not shown).

\subsection{Baroclinic eddies}

We investigate the model's ability to generate baroclinic eddies with the eddying channel test case of \cite{ilicak2012}.
This test case is an idealization of the Antarctic Circumpolar Current, the domain spanning 500 \unit{km} and 160 \unit{km} in the meridional and zonal directions, respectively.
The domain is 1000 \unit{m} deep.
At the zonal boundaries, periodic boundary conditions are applied; northern and southern boundaries are closed.
The Coriolis parameter is taken as a constant $1.2\times10^{-4}$ \unit{s^{-1}}.

\begin{figure}[th]
\centering
\includegraphics[width=0.6\textwidth]{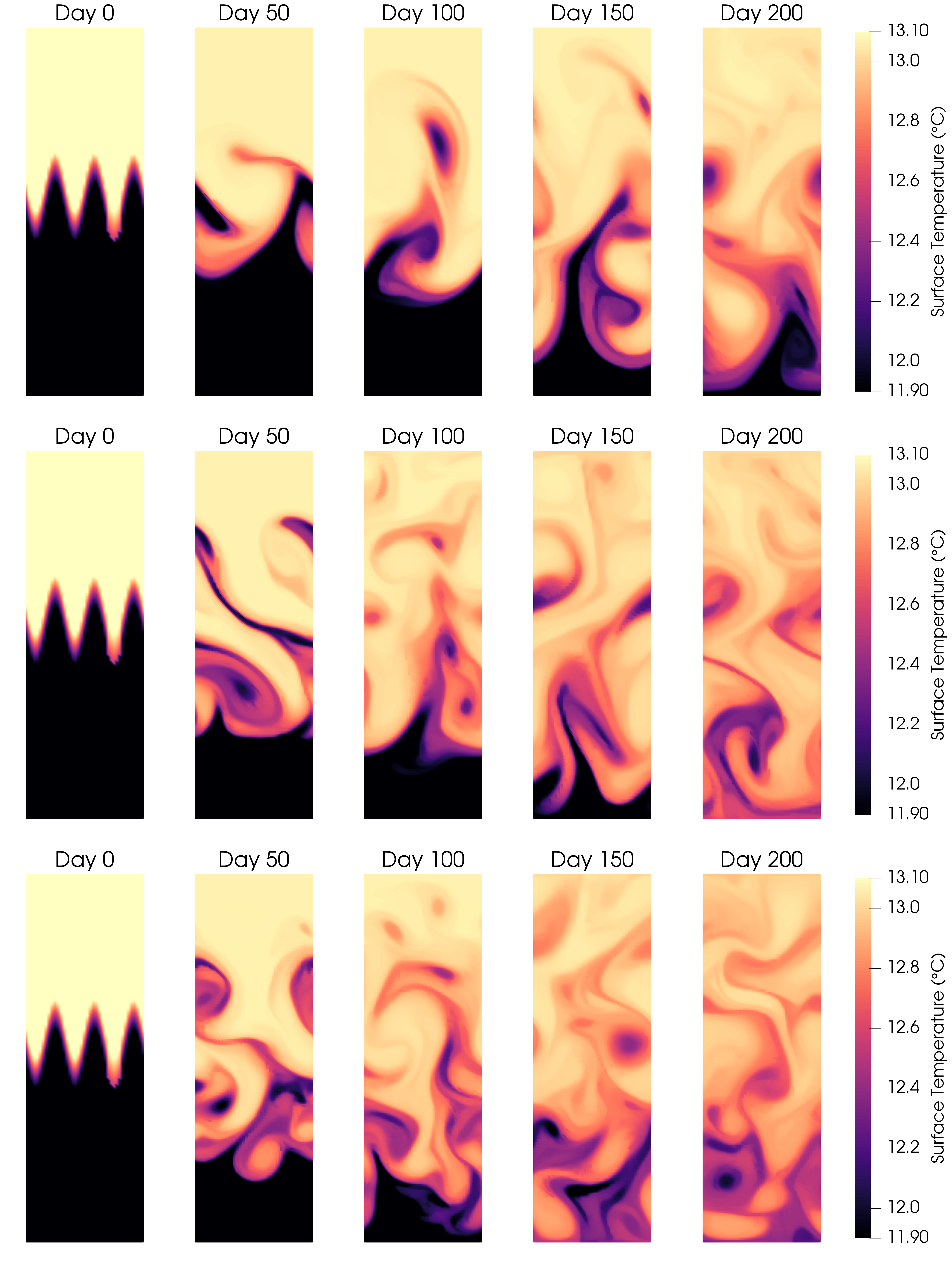}
\caption{
Sea surface temperature fields for the eddying channel test case at 4 \unit{km} horizontal mesh resolution. Horizontal viscosity is $200$ (top row), $50$ (middle), and $20$ \unit{m^2\ s^{-1}} (bottom).
These values correspond to mesh Reynolds numbers 2, 8, and 20, respectively.
}
\label{fig:eddies_sst_array}
\end{figure}

Initially, the domain is linearly stratified with warmer water at the surface.
In addition, the northern half of the domain is warmer, with a narrow sinusoidal transition band separating the warm (northern) and cold (southern) water masses
(Figure \ref{fig:eddies_sst_array}; see \citealt{petersen2015} for the definition of the initial temperature field).
Water temperature ranges between $10$ and $20$ \unit{^\circ\text{C}}.
A linear equation of state is used with $\rho_0 = 1000$ \unit{kg\ m^{-3}}, $\alpha_T=0.2$ \unit{kg\ m^{-3}\ ^\circ C^{-1}} and $T_0 = 5$ \unit{^\circ C}.
Salinity is kept at constant $35$ \unit{psu} and it does not affect density ($\beta_S=0$).
Bottom friction is parametrized by a constant drag coefficient $C_D = 0.01$.

\begin{table}[ht!]
\begin{center}
\begin{tabular}{rrrrr}
$\Delta x$ (km) & $nz$ & $\Delta t$ ($s$) & $\nu_h$ (\unit{m^2\ s^{-1}}) & $\text{Re}_h$ \\ \hline
10 & 26 & 348.39 &  10.0 & 100 \\
10 & 26 & 348.39 &  20.0 & 50 \\
10 & 26 & 348.39 &  50.0 & 20 \\
10 & 26 & 348.39 & 125.0 &  8 \\
10 & 26 & 348.39 & 200.0 &  5 \\
10 & 26 & 348.39 & 500.0 &  2 \\
4  & 40 & 140.26 &   4.0 & 100 \\
4  & 40 & 140.26 &   8.0 & 50 \\
4  & 40 & 140.26 &  20.0 & 20 \\
4  & 40 & 140.26 &  50.0 &  8 \\
4  & 40 & 140.26 & 200.0 &  2 \\
\end{tabular}
\end{center}
\caption{
Experimental setup for baroclinic eddy test case.
Listed are the horizontal mesh resolution (min. triangle edge length), number of vertical levels, time step, horizontal viscosity, and the approximate grid Reynolds number.
}\label{tab:eddies_setup}
\end{table}

The baroclinic Rossby radius of deformation is 20 \unit{km} \citep{ilicak2012}.
Horizontal mesh resolution is constant in space.
We use a regular split-quad mesh with two different mesh resolutions: eddy-permitting 10 \unit{km} and a finer 4 \unit{km} resolution.
In the vertical direction, 26 and 40 equidistant sigma levels are used in the two cases, respectively.
Simulations are carried out with different values of horizontal viscosity, the grid Reynolds number ranging from 2 to 100.
The different setups are summarized in Table \ref{tab:eddies_setup}.
Vertical viscosity is set to a constant $10^{-4}$ \unit{m^2\ s^{-1}}.

As the simulation progresses, baroclinic eddies develop at the center of the domain, quickly propagating elsewhere.
This is a spin-down experiment, i.e. the domain is a closed system with no forcing at the boundaries.
Therefore all the energy in the system originates from the initial potential energy, which is being dissipated during the simulation;
again the RPE is used as a metric for the energy transfer, or, the loss of energy due to mixing.

Figure \ref{fig:eddies_sst_array} shows the surface temperature fields at various time intervals up to 200 days after the initialization for different values of horizontal viscosity.
As expected, the model captures more mesoscale features as viscosity is decreased.
Qualitatively, the results are in agreement with ROMS and MITgcm results \citep{ilicak2012}, as well as MPAS-Ocean \citep{petersen2015}, all of which use a comparable Laplacian scheme for horizontal viscosity.

\begin{figure}[th]
\centering
\includegraphics[width=0.85\textwidth]{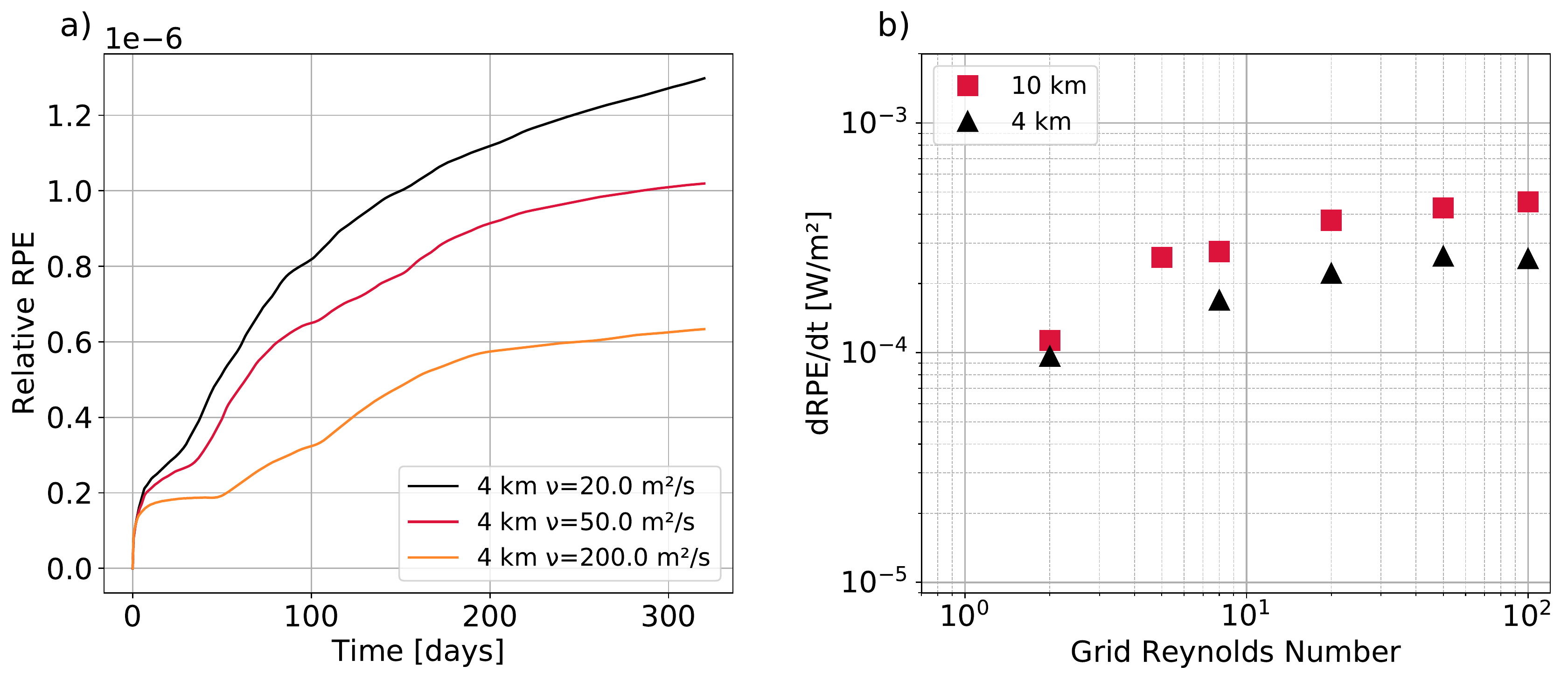}
\caption{
Diagnostics of the eddying channel test case.
(a) Evolution of normalized RPE over time in the eddying channel test case for 4 \unit{km} mesh resolution.
(b) Rate of change of RPE for different grid resolutions and grid Reynolds numbers.
The rate of change was evaluated by computing the average RPE change from day 3 to 319.
}
\label{fig:eddies_ts_rpe}
\end{figure}

The evolution of the normalized RPE during the simulation is shown in Figure \ref{fig:eddies_ts_rpe} (a) for the 4 \unit{km} mesh resolution.
The amount of mixing clearly depends on the grid Reynolds number, RPE being roughly twice as high for $\text{Re}_h=20$ compared to $\text{Re}_h=2$.
The average rate of change of RPE, averaged over days 3 to 319, is shown in Figure \ref{fig:eddies_ts_rpe} (b) for all the simulations.
As expected, the rate of change increases with larger grid Reynolds number, and with a coarser mesh.
These RPE metrics are in good agreement with results in the literature:
At $\text{Re}_h=20$ Thetis $d\text{RPE}/dt$ values are $4.3\times10^{-4}$ and $2.2\times10^{-4}$ \unit{W\ m^{-2}}, for the 10 and 4 \unit{km} resolutions.
The corresponding values for MITgcm, MOM, and POP (averaged over the days 3 to 319) are larger, at least $8\times10^{-4}$ and $3\times10^{-4}$ \unit{W\ m^{-2}}, respectively \citep[][fig. 12]{petersen2015}.	
\cite{ilicak2012} reported similar values for MITgcm and MOM.
On a hexagonal mesh, MPAS-Ocean yields smaller $d\text{RPE}/dt$ values, approximately $2\times10^{-4}$ and $7\times10^{-5}$ \unit{W\ m^{-2}} for the two resolutions, respectively \citep[values averaged over days 1--320; see fig. 12 in ][]{petersen2015}.
With a quad mesh, however, MPAS-Ocean values are approximately $2\times10^{-4}$ \unit{W\ m^{-2}} for both resolutions, therefore close to Thetis performance.

\begin{figure}[th]
\centering
\includegraphics[width=0.85\textwidth]{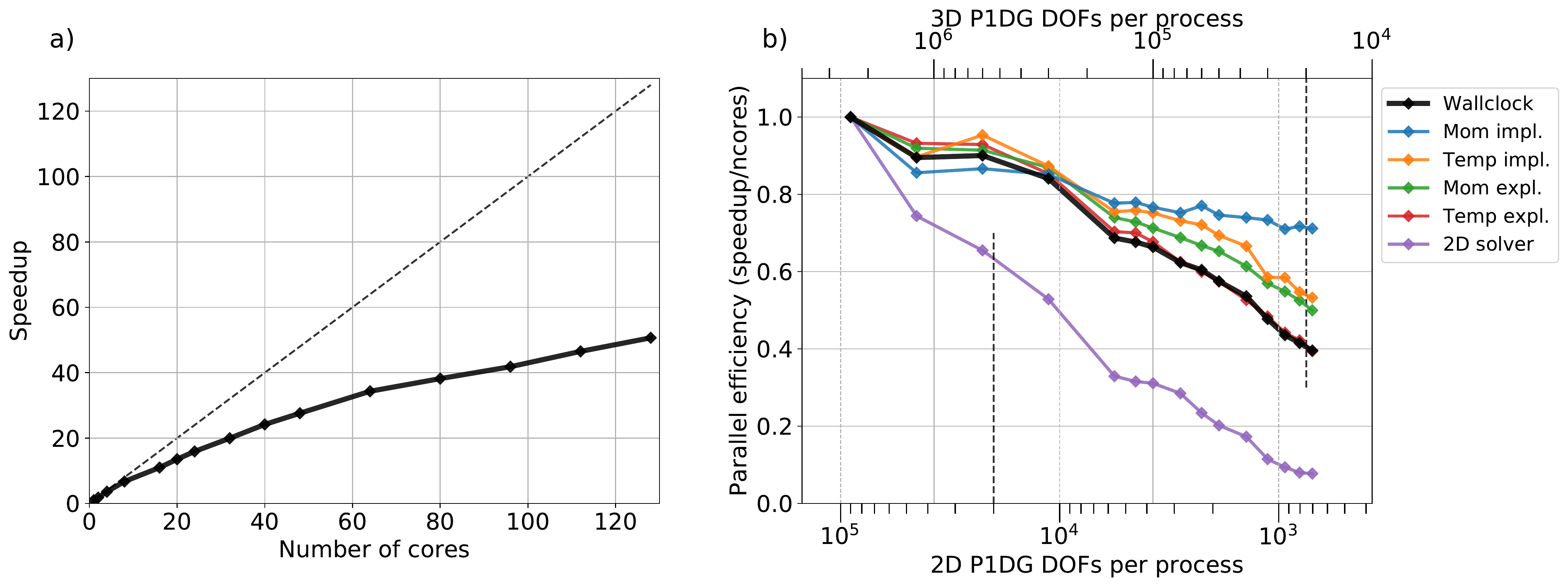}
\caption{
Strong parallel scaling for the baroclinic eddies test case on a 4 \unit{km} mesh ($\nu_h=20$ \unit{m^2\ s^{-1}}):
(a) Speedup in wallclock time versus number of processes;
(b) Parallel efficiency versus the local number of degrees of freedom (DOF) in the 3D tracer field (top axis) and the 2D $(\bar{\mathbf{u}},\eta)$ mixed system (bottom axis).
The black line is the wall clock time; colored lines stand for the time spent in different implicit or explicit solvers.
The vertical dashed lines indicate 20 000 DOFs per process for the 2D and 3D problems, respectively.
The mesh consisted of 10 000 triangles, 40 vertical levels and 400 000 prisms.
}
\label{fig:eddies_scaling}
\end{figure}

The test cases were run on a Linux cluster with 16-core Intel Xeon E5620 processors and Mellanox Infiniband interconnect.
The 320-day simulation took roughly 42 hours to run on 96 cores with the 4 \unit{km} resolution mesh and 140.26 \unit{s} time step.
It should be noted, however, that the time step employed here is smaller than the maximal allowed time step.
We also carried out a strong scaling test with the 4 \unit{km} mesh.
In the scaling test, the simulation was run for 40 time steps, recording the total elapsed wall clock time and time spent in different parts of the solver.
Figure \ref{fig:eddies_scaling} (a) shows the overall speedup up to 96 processors.
The scaling efficiency drops to roughly 50\% at 96 cores, when the local degree of freedom count for the tracer field is \mbox{25 000} (see black line Figure \ref{fig:eddies_scaling} b).
This scaling efficiency is close to typical Firedrake performance \citep{rathgeber2016}.

The scaling efficiency of the separate solvers is plotted with colored lines in Figure \ref{fig:eddies_scaling} (b).
The implicit vertical diffusion/viscosity solvers perform best due to the fact that the problem is purely local without any horizontal dependencies.
The explicit momentum solver scales almost as well, whereas the explicit tracer solver scales poorer.
The implicit 2D solver (assembly and linear solve) scales the poorest because the problem is relatively small; at 96 cores there are only around 940 degrees of freedom in the $(\bar{\mathbf{u}},\eta)$ system per core.
We have also experimented with explicit 2D solvers but they do not scale significantly better compared to the two-stage implicit scheme used herein.

To further assess the CPU cost, we compared Thetis timings against the SLIM 3D model \citep{white2008a,blaise2010,comblen2010a,karna2013} which uses a similar DG formulation but is implemented in C/C++.
The wall clock time, and parallel efficiency used by both Thetis and SLIM 3D are presented in Appendix \ref{sec:scaling_comp_slim}.
The setup, mesh, and time step were identical for the two models.
On a single core, Thetis runs 3.3$\times$ faster. On 24 cores the ratio is 4.0$\times$, and on 144 cores Thetis is still 2.2$\times$ faster than SLIM 3D.
This highlights the fact that Firedrake can deliver good parallel performance compared to models written in lower level languages.

It should be noted, however, that Thetis performance is currently not fully optimized.
We expect that the timings can be significantly improved both in terms of serial and strong scaling performance.
These will be addressed in future work.

\subsection{DOME}

\begin{figure}[th]
\centering
\includegraphics[width=0.6\textwidth]{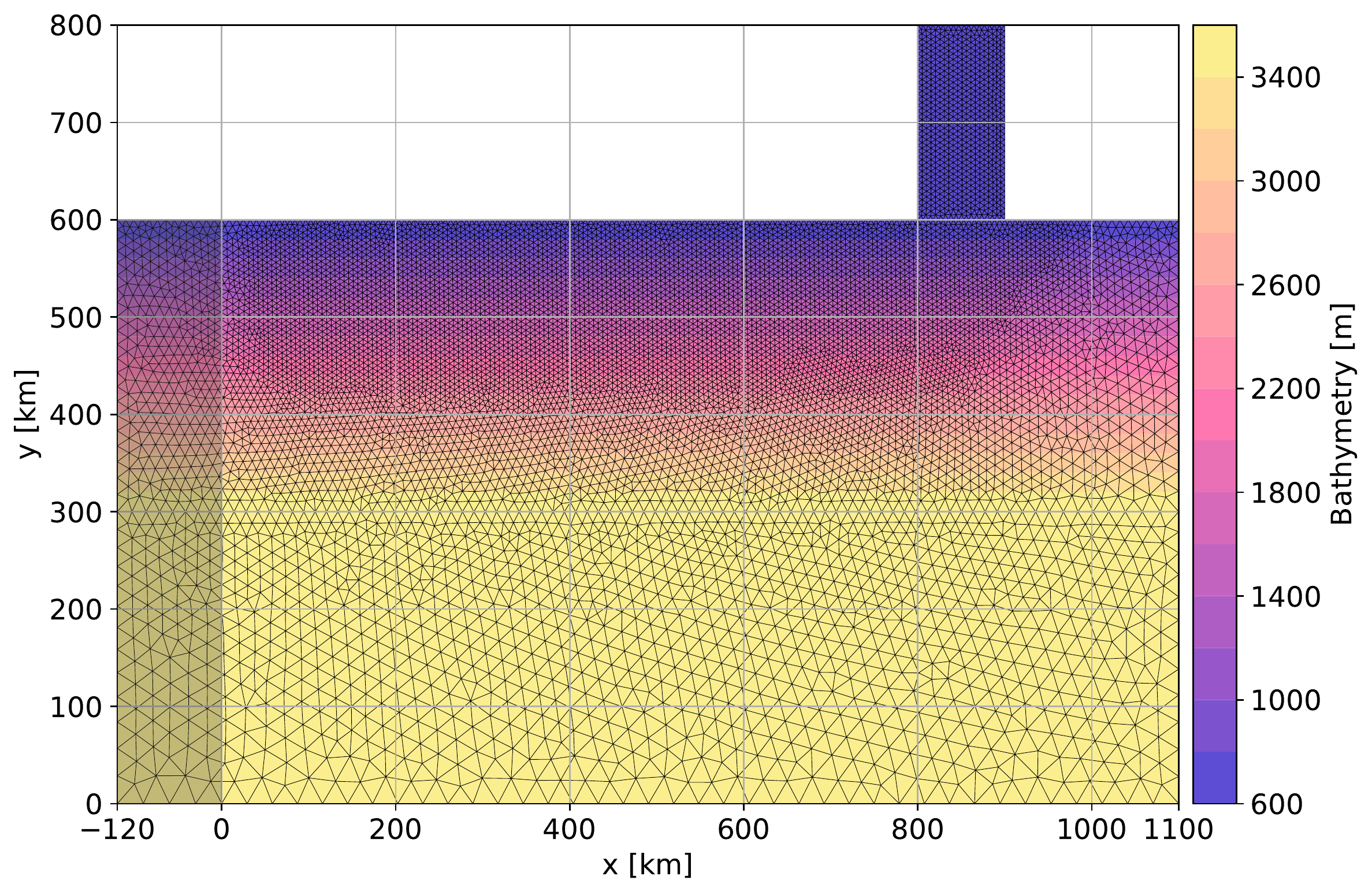}
\caption{
Horizontal mesh and bathymetry for the DOME test case.
The domain is extended 120 \unit{km} further to the west to avoid boundary effects (shaded region).
Horizontal element size ranges from 6 to 22 \unit{km}.
There are $18.8\times10^3$ triangles in the horizontal mesh and 24 uniformly distributed vertical levels resulting in $450\times10^3$ prisms and $2.7\times10^6$ tracer degrees of freedom.
}
\label{fig:dome_mesh}
\end{figure}

Next we investigate the model's ability to simulate density driven overflows with the Dynamics of Overflow Mixing and Entrainment (DOME) benchmark \citep{ezer2004,legg2006,wang2008a,burchard2008b}.
The domain is a 1100 \unit{km} by 600 \unit{km} large basin, whose depth varies linearly from 600 \unit{m} at the northern boundary to 3600 \unit{m} in the middle of the domain (see Figure \ref{fig:dome_mesh}).
To avoid boundary condition issues we have extended the domain to the west by 120 \unit{km}.
At the northern boundary, there is a 100 \unit{km} wide and 200 \unit{km} long inlet.
Initially, the basin is stably stratified with a linear temperature variation from $10$ \unit{^\circ\ C} in the deepest part of the basin to $20$ \unit{^\circ\ C} at the surface.
We use the linear equation of state
with $\rho_0 = 1000$ \unit{kg\ m^{-3}}, $\alpha_T=0.2$ \unit{kg\ m^{-3}\ {^\circ} C^{-1}} and $T_0 = 10$ \unit{^\circ C}
resulting in a $\Delta \rho=2.0$ \unit{kg\ m^{-3}} density difference.

\begin{figure}[th]
\centering
\includegraphics[width=0.6\textwidth]{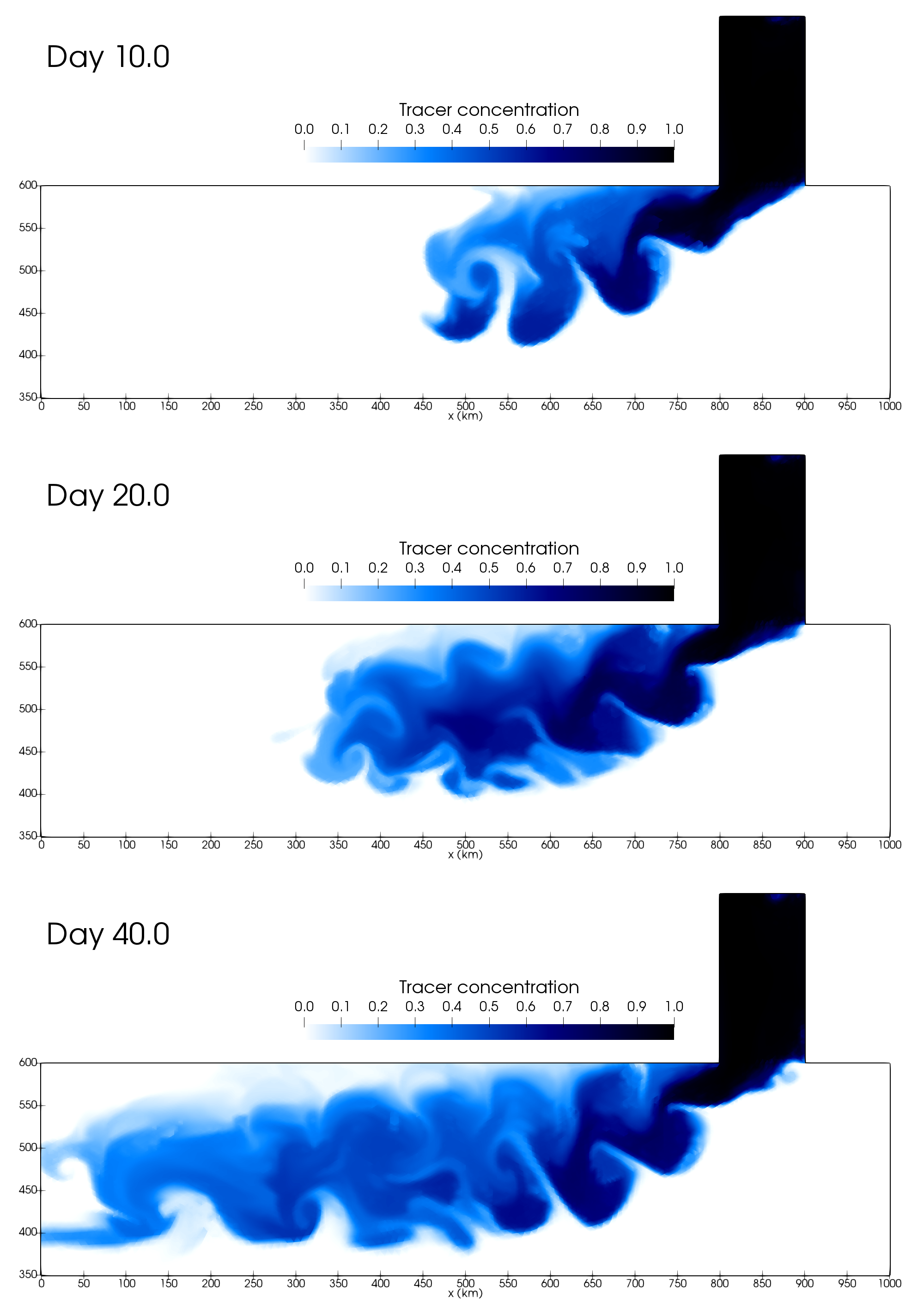}
\caption{
Bottom tracer concentration in the DOME test case after 10 (top), 20 (middle) and 40 (bottom) days.
}
\label{fig:dome_bottom_tracer}
\end{figure}

At the inlet, a dense inflow (temperature $10$ \unit{^\circ C}) is prescribed in the bottom layer, with the surface layer being at $20$ \unit{^\circ C}.
The inflow is in geostrophic balance, the thickness of the bottom layer being roughly 300 \unit{m} in the eastern end of the boundary diminishing exponentially westward \citep{legg2006}.
The total inflow in the bottom layer is $5$ Sv ($5\times10^{6}$ \unit{m^3\ s^{-1}}), the surface layer being static.
During the simulation, the fate of the inflowing waters is tracked with a passive tracer that is initially zero in the basin and unity at the inlet.
Initially, the tracer field is set to the inflow conditions in the northern part of the basin ($y>650$ \unit{km}). Velocity is set to zero everywhere. 
The eastern and southern boundaries of the basin are closed.
The western boundary is open with radiation boundary conditions, and a 100 \unit{km} wide band where the temperature is relaxed to the initial condition.

The domain is discretized with an unstructured grid (Figure \ref{fig:dome_mesh}).
Horizontal mesh resolution is 6 \unit{km} near the northern boundary, increasing southward.
24 vertical sigma levels are used.
Over the slope, the mesh resolution was designed to result in a hydrostatic consistency metric $r<1.5$ \citep{beckmann1993}.
Horizontal viscosity is set to a constant $50$ \unit{m^2\ s^{-1}}, which corresponds to $\text{Re}_h\approx200$ at the inlet.
Horizontal diffusivity is constant at $10$ \unit{m^2\ s^{-1}}.
Vertical viscosity and diffusivity are parametrized by the Pacanowski-Philander scheme as described in Section \ref{sec:turbulence}.
Bottom friction is parametrized with a quadratic drag coefficient $C_d=2\times10^{-3}$ \citep{legg2006,wang2008a}.
A quadratic function space is used for the baroclinic head and internal pressure gradient as discussed in Section \ref{sec:weak_int_pg}.

As the inflowing current reaches the basin, it turns to the west and forms a coastal plume that is approximately 150 \unit{km} wide (Figure \ref{fig:dome_bottom_tracer}).
The plume detaches from the lateral boundary as it flows westward and along the bottom slope.
As the dense water mass meets the stratified ocean, the plume becomes unstable and starts to generate eddies and internal waves.
The most vigorous eddies are found in the first 300 \unit{km} after the inlet ($x=500\text{--}800$ \unit{km}), after which the plume is more mixed and quiescent.
Overall the plume is shallow; most of the passive tracer is concentrated within 200 \unit{m} of the bottom.
Qualitatively, the extent and propagation of the plume, and its eddy structure are in good agreement with the literature \citep[e.g.,][]{burchard2008b,wang2008a}.
The results show that Thetis is able to represent eddying flows over sloping bathymetry, generating and maintaining strong gradients between water masses.
The sharpest fronts in the simulation encompass only one or two elements.

\begin{figure}[th]
\centering
\includegraphics[width=0.6\textwidth]{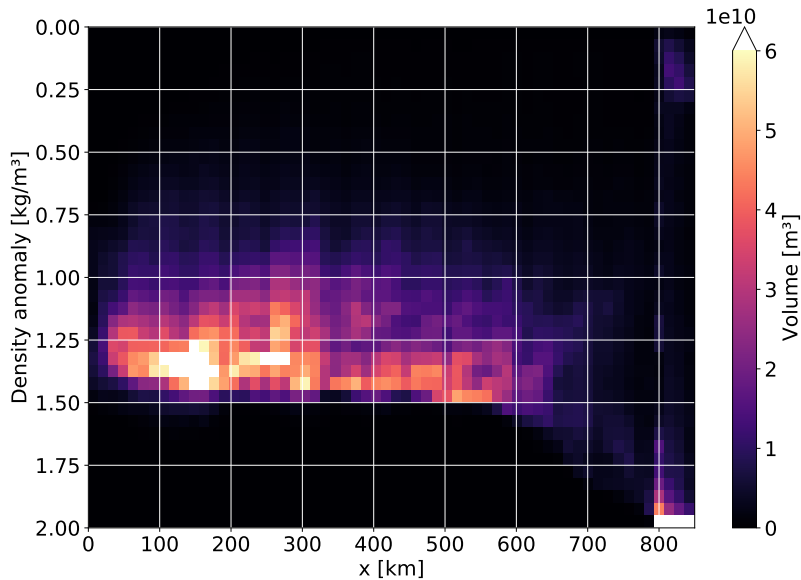}
\caption{
Histogram of tracer in the DOME test case versus the $x$ coordinate and density class.
At the mouth of the inlet ($x = 800$ \unit{km}) the inflowing waters are dense; they get entrained higher up in the density spectrum as they are being transported downstream.
The data are averaged over one week after day 40.
}
\label{fig:dome_tracer_histogram}
\end{figure}

Figure \ref{fig:dome_tracer_histogram} shows the distribution of the inflowing tracer concentration as a function of water density and the $x$-axis.
The inflowing waters are initially very dense but get mixed to lower density as the plume advances along the coast.
The histogram shows that the plume volume is low in the first 150 \unit{km} after the inlet ($x=650\text{--}800$ \unit{km}) where the plume accelerates.
After $x=650$ \unit{km} the plume slows down and starts to accumulate in volume.
The density of the main plume occupies ranges from $0.8$ to $1.5$ \unit{kg\ m^{-3}}, the peak being around $1.28$ \unit{kg\ m^{-3}}.
The rate of entrainment can be used as a metric for mixing.
Results herein are similar to those presented in literature:
\cite{wang2008a} present a mean density anomaly of $1.5$ \unit{kg\ m^{-3}} for their terrain following FESOM model configuration.

The 47-day simulation took roughly 42 hours to run on 90 cores with a 39.65 \unit{s} time step on the same Linux cluster.

\conclusions
\label{sec:discussion}

This paper describes a DG implementation of an eddy-permitting, unstructured grid coastal ocean model.
The solver is second-order accurate in space and time.
We have demonstrated that the formulation is fully conservative and preserves monotonicity.
The test cases indicate that the model is capable of reproducing the expected physical behavior, including baroclinic eddies.
Moreover, numerical mixing is well-controlled and comparable to other established structured grid models, such as MITgcm and ROMS, and the large-scale finite volume model MPAS-Ocean.
Finding an accurate formulation is important as commonly-used unstructured grid models tend to be overly diffusive, preventing accurate modeling of certain coastal domains \citep[e.g.,][]{karna2015}.
The formulation presented herein thus contributes to the development of more accurate next-generation coastal ocean models.

Future work will include solving the equations on a sphere, DG implementation of a biharmonic viscosity operator, two-equation turbulence closure models, wetting-drying treatment, development of an adjoint solver, as well as improving the computational efficiency and parallel scaling of the solver.

\section{Code availability}

All code used to perform the experiments in this papers is publicly
available.  Firedrake, and its components, may be obtained from
\url{www.firedrakeproject.org}; Thetis from \url{thetisproject.org}.

For reproducibility, we also cite archives of the exact software
versions used to produce the results in this paper.  All major
Firedrake components have been archived on Zenodo
\citep{zenodo/Firedrake}.  This record collates DOIs for the
components, and can be installed following the instructions at
\url{www.firedrakeproject.org/download.html} with \texttt{firedrake-install
  --doi 10.5281/zenodo.1407898}.  Thetis itself has been archived at
\citep{zenodo/Thetis}.

\section{Data availability}

No external data were used in this manuscript.

\appendix
\section{Source terms for the baroclinic MMS test}\label{sec:baroclinic_mms_source_terms}

Using the analytical velocity and temperature fields we can derive the steady state solution for the remaining fields

\begin{align}
\eta_a &= 0, \\
\bar{u}_a &= \frac{1}{6} \sin{\left (3 \right )} \sin{\left (\frac{2 \pi}{L_{x}} x \right )}, \\
\bar{v}_a &= \frac{1}{3} \sin{\left (\frac{z}{2 h} \right )} \cos{\left (\frac{\pi y}{L_{y}} \right )}, \\
u_a' &= u_a - \bar{u}_a, \\
v_a' &= v_a - \bar{v}_a,
\end{align}
\begin{align}
w_a &= \frac{\pi h}{3 L_{x} L_{y}} \left(2 L_{x} \left(- \cos{\left (\frac{z}{2 h} \right )} + \cos{\left (\frac{1}{2} \right )}\right) \sin{\left (\frac{\pi y}{L_{y}} \right )} - L_{y} \left(\sin{\left (\frac{3 z}{h} \right )} + \sin{\left (3 \right )}\right) \cos{\left (\frac{2 \pi}{L_{x}} x \right )}\right), \\
r_a &= \frac{\alpha_T}{\rho_{0}} \left(T_{0} z - 15 h \sin{\left (\frac{z}{h} \right )} \sin{\left (\frac{\pi x}{L_{x}} \right )} \sin{\left (\frac{\pi y}{L_{y}} \right )} - 15 z\right).
\end{align}

Now we can evaluate the different terms that appear in the momentum and tracer equations:

\begin{align}
f(\mathbf{e}_z\wedge\bar{\mathbf{u}})_x &= \frac{2 f_{0}}{3} \left(- \cos{\left (\frac{1}{2} \right )} + 1\right) \cos{\left (\frac{\pi y}{L_{y}} \right )}, \\
f(\mathbf{e}_z\wedge\bar{\mathbf{u}})_y &= \frac{f_{0}}{6} \sin{\left (3 \right )} \sin{\left (\frac{2 \pi}{L_{x}} x \right )}, \\
\boldsymbol{\nabla}_h\cdot\left(H\bar{\mathbf{u}}\right) &= \frac{\pi h}{3 L_{x} L_{y}} \left(2 L_{x} \left(- \cos{\left (\frac{1}{2} \right )} + 1\right) \sin{\left (\frac{\pi y}{L_{y}} \right )} + L_{y} \sin{\left (3 \right )} \cos{\left (\frac{2 \pi}{L_{x}} x \right )}\right),
\end{align}

\begin{align}
(\mathbf{F}_{\text{pg}})_{x} &= \frac{15 \pi \alpha_T h}{L_{x} \rho_{0}} g \sin{\left (\frac{z}{h} \right )} \sin{\left (\frac{\pi y}{L_{y}} \right )} \cos{\left (\frac{\pi x}{L_{x}} \right )}, \\
(\mathbf{F}_{\text{pg}})_{y} &= \frac{15 \pi \alpha_T h}{L_{y} \rho_{0}} g \sin{\left (\frac{z}{h} \right )} \sin{\left (\frac{\pi x}{L_{x}} \right )} \cos{\left (\frac{\pi y}{L_{y}} \right )}, \\
(\boldsymbol{\nabla}_h \cdot (\mathbf{u} \mathbf{u}))_{x} &= \frac{\pi}{2 L_{x}} \sin{\left (\frac{2 \pi}{L_{x}} x \right )} \cos^{2}{\left (\frac{3 z}{h} \right )} \cos{\left (\frac{2 \pi}{L_{x}} x \right )}, \\
(\boldsymbol{\nabla}_h \cdot (\mathbf{u} \mathbf{u}))_{y} &= - \frac{\pi}{9 L_{y}} \sin^{2}{\left (\frac{z}{2 h} \right )} \sin{\left (\frac{\pi y}{L_{y}} \right )} \cos{\left (\frac{\pi y}{L_{y}} \right )}, \\
\frac{\partial \left(w u \right)}{\partial z} &= \frac{\pi \sin{\left (\frac{3 z}{h} \right )}}{2 L_{x} L_{y}} \left(2 L_{x} \left(\cos{\left (\frac{z}{2 h} \right )} - \cos{\left (\frac{1}{2} \right )}\right) \sin{\left (\frac{\pi y}{L_{y}} \right )} + L_{y} \left(\sin{\left (\frac{3 z}{h} \right )} + \sin{\left (3 \right )}\right) \cos{\left (\frac{2 \pi}{L_{x}} x \right )}\right) \sin{\left (\frac{2 \pi}{L_{x}} x \right )}, \\
\frac{\partial \left(w v \right)}{\partial z} &= - \frac{\pi \cos{\left (\frac{z}{2 h} \right )}}{18 L_{x} L_{y}} \left(2 L_{x} \left(\cos{\left (\frac{z}{2 h} \right )} - \cos{\left (\frac{1}{2} \right )}\right) \sin{\left (\frac{\pi y}{L_{y}} \right )} + L_{y} \left(\sin{\left (\frac{3 z}{h} \right )} + \sin{\left (3 \right )}\right) \cos{\left (\frac{2 \pi}{L_{x}} x \right )}\right) \cos{\left (\frac{\pi y}{L_{y}} \right )}, \\
f(\mathbf{e}_z\wedge \mathbf{u}')_x &= \frac{f_{0}}{3} \left(- \sin{\left (\frac{z}{2 h} \right )} - 2 + 2 \cos{\left (\frac{1}{2} \right )}\right) \cos{\left (\frac{\pi y}{L_{y}} \right )}, \\
f(\mathbf{e}_z\wedge \mathbf{u}')_y &= \frac{f_{0}}{6} \left(3 \cos{\left (\frac{3 z}{h} \right )} - \sin{\left (3 \right )}\right) \sin{\left (\frac{2 \pi}{L_{x}} x \right )},
\end{align}

\begin{align}
\boldsymbol{\nabla}_h \cdot (\mathbf{u} T) &= \frac{5 \pi}{L_{x} L_{y}} \left(L_{x} \sin{\left (\frac{z}{2 h} \right )} \cos^{2}{\left (\frac{\pi y}{L_{y}} \right )} + 3 L_{y} \sin{\left (\frac{\pi y}{L_{y}} \right )} \cos{\left (\frac{3 z}{h} \right )} \cos^{2}{\left (\frac{\pi x}{L_{x}} \right )}\right) \sin{\left (\frac{\pi x}{L_{x}} \right )} \cos{\left (\frac{z}{h} \right )}, \\
\frac{\partial \left(w T \right)}{\partial z} &= \frac{5 \pi}{L_{x} L_{y}} \bigg[ 2 L_{x} \left(\cos{\left (\frac{z}{2 h} \right )} - \cos{\left (\frac{1}{2} \right )}\right) \sin{\left (\frac{\pi y}{L_{y}} \right )} + \nonumber \\
 & \quad\quad\quad\quad\ L_{y} \left(\sin{\left (\frac{3 z}{h} \right )} + \sin{\left (3 \right )}\right) \cos{\left (\frac{2 \pi}{L_{x}} x \right )} \bigg]     \sin{\left (\frac{z}{h} \right )} \sin{\left (\frac{\pi x}{L_{x}} \right )} \sin{\left (\frac{\pi y}{L_{y}} \right )}.
\end{align}

These terms are added as source terms to the right hand side of the equations \eqref{eq:mom2d}, \eqref{eq:freesurface}, \eqref{eq:mom3d}, and \eqref{eq:tracer}.
In the weak form this corresponds to multiplying the analytical function by the test function and integrating over the domain.
The solutions were derived using the SymPy symbolic mathematics Python library \citep{sympy2017}.

\section{CPU cost comparison against SLIM}\label{sec:scaling_comp_slim}

A strong scaling test was carried out with both Thetis and SLIM 3D model \citep{white2008a,blaise2010,comblen2010a,karna2013} using the baroclinic eddies test case.
These tests were carried out on a Linux cluster with 16-core Intel Xeon E5620 processors and Mellanox Infiniband interconnect.
The total time spent to run 40 time steps is presented in Table \ref{tab:scaling_timings}.
The table also lists the speed-up $s_i = T_0/T_i$, where $T_i$ stands for the wall clock time for $i$ cores, and the parallel efficiency $p = s_i/i$.
For an ideal model the parallel efficiency remains at unity.
The results show that on a single core Thetis runs approximately 3.3$\times$ faster than SLIM. On 24 cores the ratio is 4.0$\times$, and on 144 cores Thetis is still 2.2$\times$ faster.

\begin{table}[ht!]
\begin{center}
\begin{tabular}{|r|rr|r|rr|rr|}
\hline
Nb cores   & \multicolumn{2}{|c|}{Wallclock (\unit{s})} & Ratio & \multicolumn{2}{|c|}{Speed-up} & \multicolumn{2}{|c|}{Efficiency} \\ \hline
           & Thetis  & SLIM                   & $\frac{T_{SLIM}}{T_{Thetis}} $ & Thetis & SLIM                  & Thetis & SLIM \\ \hline
1          & 1778.71 & 5928.32                & 3.33 &  1.00 &  1.00                 &   1.00 &  1.00 \\
2          & 1034.64 & 4802.34                & 4.64 &  1.72 &  1.23                 &   0.86 &  0.62 \\
4          &  500.11 & 2380.74                & 4.76 &  3.56 &  2.49                 &   0.89 &  0.62 \\
8          &  290.61 & 1284.08                & 4.42 &  6.12 &  4.62                 &   0.77 &  0.58 \\
16         &  206.97 &  675.14                & 3.26 &  8.59 &  8.78                 &   0.54 &  0.55 \\
20         &  141.17 &  524.61                & 3.72 & 12.60 & 11.30                 &   0.63 &  0.57 \\
24         &  110.83 &  440.09                & 3.97 & 16.05 & 13.47                 &   0.67 &  0.56 \\
32         &   88.03 &  330.00                & 3.75 & 20.21 & 17.96                 &   0.63 &  0.56 \\
40         &   73.17 &  260.47                & 3.56 & 24.31 & 22.76                 &   0.61 &  0.57 \\
48         &   64.16 &  222.79                & 3.47 & 27.72 & 26.61                 &   0.58 &  0.55 \\
64         &   56.62 &  158.31                & 2.80 & 31.41 & 37.45                 &   0.49 &  0.59 \\
80         &   49.48 &  127.95                & 2.59 & 35.95 & 46.33                 &   0.45 &  0.58 \\
96         &   43.64 &  109.10                & 2.50 & 40.76 & 54.34                 &   0.42 &  0.57 \\
112        &   39.68 &   95.24                & 2.40 & 44.83 & 62.25                 &   0.40 &  0.56 \\
128        &   36.91 &   83.37                & 2.26 & 48.19 & 71.11                 &   0.38 &  0.56 \\
144        &   35.76 &   78.05                & 2.18 & 49.74 & 75.96                 &   0.35 &  0.53 \\
\hline
\end{tabular}
\end{center}
\caption{
CPU time in the baroclinic eddies test case for Thetis and SLIM model.
Both models were ran on identical triangular mesh (4 \unit{km} resolution, 40 vertical levels) using $\nu = 20$ \unit{m^2\ s^{-1}}, and 140 \unit{s} time step. The wall clock time was recorded over 40 iterations.
}\label{tab:scaling_timings}
\end{table}




\authorcontribution{
Tuomas K\"{a}rn\"{a} designed and implemented most of the solver and carried out the numerical simulations.
Stephan Kramer and Lawrence Mitchell contributed to the design and implementation of the model.
Ant\'onio Baptista, David Ham and Matthew Piggott supervised the work and guided the implementation of the model and the manuscript.
}



\begin{acknowledgements}
The National Science Foundation partially supported this research through
cooperative agreement OCE-0424602. The National Oceanic and Atmospheric
Administration (NA11NOS0120036 and AB-133F-12-SE-2046), Bonneville Power
Administration (00062251) and Corps of Engineers (W9127N-12-2-007 and
G13PX01212) provided partial motivation and additional support. This
work was supported by the UK's Engineering and Physical Science Research
Council [grant numbers EP/M011054/1, EP/L000407/1]; and the Natural
Environment Research Council [grant number NE/K008951/1]. This work used
the Extreme Science and Engineering Discovery Environment (XSEDE), which is
supported by National Science Foundation grant number ACI-1053575. The authors
acknowledge the Texas Advanced Computing Center (TACC) at The University of
Texas at Austin for providing HPC resources that have contributed to the
research results reported within this paper.
\end{acknowledgements}



%
%
%

\bibliographystyle{copernicus}
\bibliography{references,zenodo}

\end{document}